\documentclass[reprint,superscriptaddress,amsmath,amssymb,aps,prb,floatfix]{revtex4-2}

\usepackage{graphicx}
\usepackage{dcolumn}
\usepackage{bm}
\usepackage{physics}
\usepackage[colorlinks,
    linkcolor=blue,
    anchorcolor=red,
    citecolor=green
]{hyperref}

\usepackage{simpler-wick}
\usepackage{array}
\usepackage{cancel}
\usepackage{times}
\usepackage[caption = false]{subfig}

\newcommand{\mi}{\mathrm{i}}
\newcommand{\me}{\mathrm{e}}

\begin{document}

\title{Topological symplectic Kondo effect}

\author{Guangjie Li}
\affiliation{%
Department of Physics and Astronomy, Purdue University, West Lafayette, Indiana 47907, USA.
}

\author{Elio J. K{\"o}nig}
\thanks{These authors contributed equally}
\affiliation{%
Max-Planck Institute for Solid State Research, 70569 Stuttgart, Germany
}%

\author{Jukka I. V{\"a}yrynen}
\thanks{These authors contributed equally}
\affiliation{%
Department of Physics and Astronomy, Purdue University, West Lafayette, Indiana 47907, USA.
}

\newcommand{\PRLSection}[1]{\textcolor{blue}{\textit{#1}}}

\date{\today}

\begin{abstract}
Multiple conduction channels interacting with a quantum impurity -- a spin in the conventional ``multi-channel Kondo effect'' or a topological mesoscopic device (``topological Kondo effect'') -- has been proposed as a platform to realize anyonic quasi-particles. 
However, the above implementations require either perfect channel symmetry or the use of Majorana fermions. Here we propose a Majorana-free mesoscopic setup which implements the Kondo effect of the symplectic Lie group and can harbor emergent anyons (including Majorana fermions, Fibonacci anyons, and $\mathbb Z_3$ parafermions) even in the absence of perfect channel symmetry. In addition to the detailed prescription of the implementation, we present the strong coupling solution by mapping the model to the multi-channel Kondo effect associated to an internal $SU(2)$  symmetry and exploit conformal field theory (CFT) to predict the non-trivial scaling of a variety of observables, including conductance, as a function of temperature. 
This work does not only open the door for robust Kondo-based anyon platforms,
but also sheds light on the physics of strongly correlated materials with competing order parameters.

\end{abstract}

\maketitle

\PRLSection{Introduction.} 
The realization of fault tolerant quantum computation is a major goal of present day quantum research. Amongst the various hardware platforms suitable for this application, topologically ordered states with anyonic excitations are particularly appealing~\cite{NayakDasSarma2008}, as the robustness against noise and errors is a fundamental, intrinsic property of these quantum many-body phases. A classic platform for realizing anyons which has gained renewed interest in %
mesoscopic systems 
are frustrated and overscreened Kondo impurity models~\cite{2007Natur.446..167P, 2018Sci...360.1315I}.   

\begin{figure}[tb]
\centering
\includegraphics[scale=0.92]{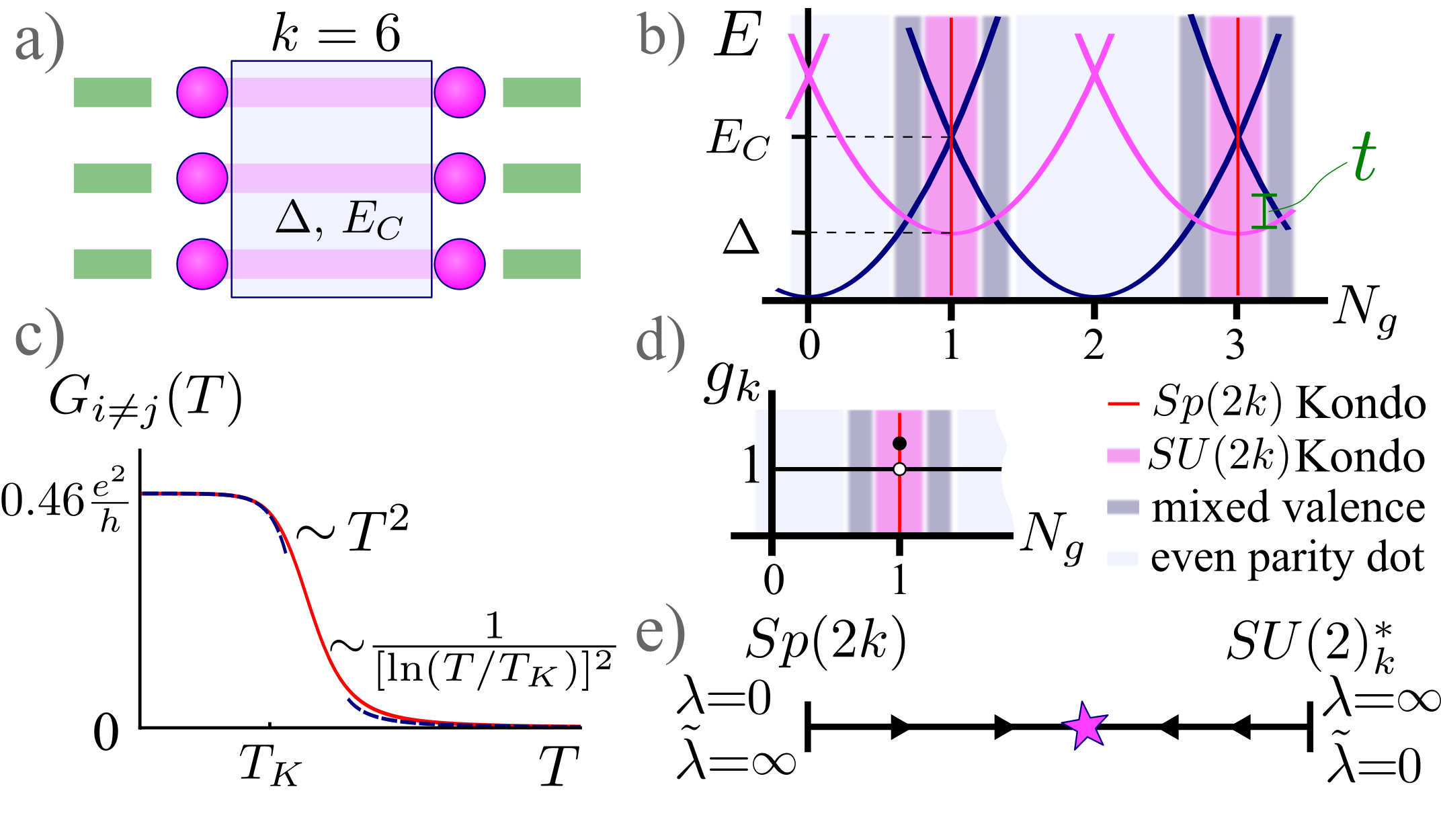}
\caption{a) Schematics of proposed implementation for $k=6$ with $\Delta$ the proximity-induced gap and $E_C$ the charging energy. Light green squares are the leads and light purple dots are the ends of 1D topological systems. The light gray square is a superconducting island.  b) Energy levels as a function of gate voltage (as imposed by charge $N_g$;  we take $\Delta = 0.4E_C$ ). The tunneling strength $t$ is between the dots and the leads (nearest sites). Dark blue (light purple) curves correspond to states with even (odd) fermion parity. The background shading corresponds to different effective low-energy theories (see legend). c) Transconductance. The solid red line interpolates between the low- and high-temperature asymptotic behavior. The conductance quantization at $T=0$ is universal, cf.~Eq.~\eqref{eq:G}, and equal to $(4/3)\sin^2(\pi/5) \approx 0.46e^2/h$ for $k=3$. In the weak coupling regime, the conductance has a logarithmic temperature-dependence.  d) Ground state degeneracy. e) Schematic RG flow illustrating the duality between $Sp(2k)$ Kondo effect and $k$-channel $SU(2)$ Kondo effect at spin $S = (k-1)/2$. }
\label{fig:SetupStates}
\end{figure}

The $SU(2)$ Kondo effect is a paradigmatic model of quantum many-body physics~\cite{2007Natur.446..167P,2015Natur.526..233I,KellerGoldhaberGordon2015,2018Sci...360.1315I, PouseGoldhaberGordon2022,PhysRevLett.130.146201} which merges the physics of strong electronic correlations and entanglement, whilst its strong coupling physics is still amenable to non-perturbative analytical methods such as Bethe \textit{ansatz}~\cite{Andrei1980,Wiegman1980,TsvelickWiegmann1983Review}, CFT~\cite{Tsvelick1990, affleck1991critical}, and Abelian bosonization~\cite{EmeryKivelson1992}. Even though the impurity spin in the conventional Kondo effect is perfectly screened at strong coupling, the overscreened multi-channel Kondo (MCK) effect, in which $k>2S$ electronic baths compete for screening a single spin-$S$, is one of the earliest examples of quantum criticality and local non-Fermi liquid (FL) behavior, and harbors a remnant zero temperature impurity entropy~\cite{AndreiDestri1984,TsvelickWiegmann1985,Tsvelick1985} $S_{\mathrm{imp}} = \ln(g_k)$ with $g_k = \sqrt{2},(1 + \sqrt{5})/2, \sqrt{3}, \dots$  for $S=1/2$ and $k = 2,3,4, \dots$ consistent with the quantum dimensions of Ising, Fibonacci, and $\mathbb Z_3$ parafermionic anyons. It has thus recently been proposed to exploit these anyons for quantum information theoretical applications~\cite{LopesSela2020,Komijani2020,DorSela2022,Lotem22,Lotem23}, but a major technical difficulty is that multichannel Kondo physics, even with $SU(N)$ and $N>2$, is unstable with respect to unequal coupling to different electronic baths.

Stable overscreened fixed points may be achieved by using strongly interacting~\cite{FieteNayak2008,KoenigTsvelik2020} or higher spin~\cite{PhysRevB.54.14918} conduction electrons, or by going beyond the conventional $SU(2)$ group. A recent example of the latter is the orthogonal Kondo  effect in which spin-polarized conduction electrons couple to an impurity spin transforming under the  group $SO(M)$. The orthogonal Kondo effect for arbitrary $M$ can be realized with the use of Majorana Cooper pair boxes~\cite{PhysRevLett.109.156803,PhysRevLett.110.216803,PhysRevLett.113.076401,PhysRevLett.113.076404}, in which case it is called the topological Kondo  effect. While fascinating, this implementation is temporarily elusive as the control over Majorana devices is still developing. Another, Majorana-free implementation for the special case $M = 5$ was recently proposed~\cite{MitchellAffleck2021,LibermanSela2021} and it was argued that Ising anyons (Majorana) are emergent at the infrared.

In this paper, we propose a mesoscopic setup, see Fig.~\ref{fig:SetupStates} realizing the symplectic Kondo effect as a platform for anyons and potentially for measurement-only topological quantum computation~\cite{BondersonNayak2008}.
Following Cartan's classification of Lie groups \cite{georgi2000lie,Kimura2021}, we here explore the third remaining type of Lie group $Sp(2k)$, i.e. a symplectic Kondo Hamiltonian
\begin{equation}
    H_K=\lambda \, \sum_{A=1}^{k(2k+1)} S^A J^A  \,, \label{eq:HKondo}
\end{equation}
in which the symplectic impurity ``spin'' operators $S_A$ transform in the fundamental $2k$-dimensional representation, and $J^A=c_0^\dagger T_A c_0$ is the symplectic spin of conduction electrons; the  spinor $c_{a}=(c_{a,1,\uparrow},\cdots,c_{a,k,\uparrow},c_{a,1,\downarrow},\cdots,c_{a,k,\downarrow})^T$ for site $a$ has $2k$ components with $i = 1, \dots k$ denoting the lead index 
and $\sigma =\uparrow, \downarrow$ the physical spin. 
Despite the $2k$ components of the spinor,  Eq.~(\ref{eq:HKondo}) is still a one-channel Kondo model and will therefore not suffer from channel anisotropy.
The $2k\times 2k $ matrices $T_A = -\sigma_y T_A^T \sigma_y$ denote $Sp(2k)$ generators in the fundamental representation~\cite{footnote1}. 
We present a mesoscopic implementation of this effect for arbitrary $k$, {the phase diagram for this nano-device},  characteristic signatures in transport measurements as well as a solution of the symplectic Kondo effect in the strong coupling limit.

From the perspective of materials science, symplectic Kondo models are theoretically appealing as they allow for a proper definition of time reversal symmetry and thus for large-N descriptions of heavy fermion superconductors~\cite{FlintColeman2008}. At the same time $SO(5) \sim Sp(4)$ theories of cuprates are popular approaches to account for competing orders~\cite{DemlerZhang2004}. From the viewpoint of quantum information theory, the symplectic Kondo effect allows for the arguably most robust way of realizing anyons in impurity models: In addition to the aforementioned stable implementation of non-trivial anyons, earlier work using CFT~\cite{Kimura2021,MitchellAffleck2021,LibermanSela2021} demonstrates that -- contrary to standard multi-channel Kondo phenomenology --  the leads behave FL-like (suggesting relatively strong decoupling of anyons and conduction electrons) and that Fibonacci anyons (which are the simplest anyons allowing for universal quantum computation) can not be realized in the simplest realization of the topological Kondo effect, but are accessible in the present $Sp(6)$ setup.

\PRLSection{Implementation of the $Sp(2k)$ Kondo model.}
We consider $k$ spinful fermionic zero-energy states coupled to a floating s-wave superconductor, see Fig.~\ref{fig:SetupStates}a.  These states may stem from a
time-reversal symmetric higher-order topological insulator, resonant levels of quantum dots, or a set of Su-Schrieffer-Heeger chains. %
The  low-energy Hamiltonian of our topological quantum dot is, 
\begin{align}
H_{\text{d}} =& E_C (2 \hat{N}_C + \hat n_d - N_g)^2 \nonumber \\
   & - \frac{1}{2} \Delta \sum_{i = 1}^k \sum_{\sigma \sigma'}\me^{-\mi \phi} d^\dagger_{i,\sigma} (\sigma_y)_{\sigma \sigma'} d^\dagger_{i, \sigma'} + \text{H.c.}, \label{eq:tunnelH}
\end{align}
where $\hat n_d = \sum_{i,\sigma} d^\dagger_{i, \sigma} d_{i \sigma}$ is the total charge in the edge states and $\hat N_C = -\mi \partial_\phi$ is the number operator of the Cooper pairs of the s-wave superconductor. The Hamiltonian Eq.~(\ref{eq:tunnelH}) 
conserves the total number of electrons $\hat N_{\rm tot} = 2 \hat{N}_C + \hat n_d $, controllable by the gate charge $N_g$. We assume that the island size exceeds the superconducting coherence length, so that crossed-Andreev reflection as well as hybridization of zero modes can be neglected, and that the proximity-induced gap $\Delta$ on the boundary states of the topological wires is less than the  bulk gap, allowing us to ignore quasiparticle states of the parent superconductor in Eq.~(\ref{eq:tunnelH}).  We also take the gap to be smaller than the charging energy,  $\Delta < E_C$, enabling a ground state with an odd number of electrons. 
We ignore additional mutual charging energies between the zero modes, which is a good assumption when the central superconducting island has a large normal-state conductivity~\cite{PhysRevB.101.125108,PhysRevB.101.241414}. 

\begin{figure}[tb]
\centering
\includegraphics[scale=0.98]{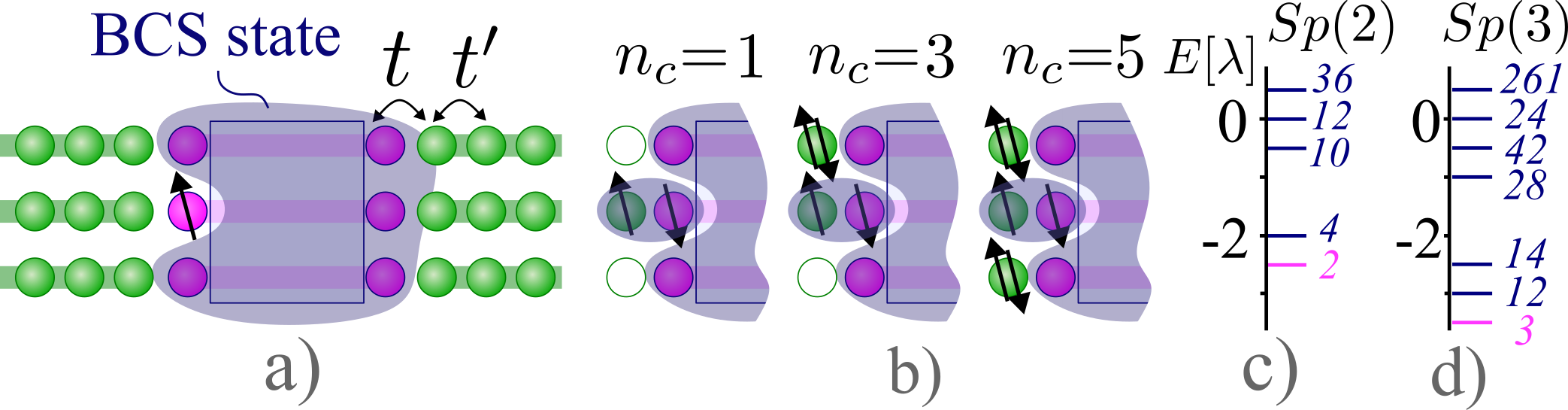}
\caption{a)
The $2k$-degenerate ground state in the odd parity sector is given by a BCS state supplemented by one unpaired electron. b) Illustration of $k$ charge degenerate ground states in the extreme strong coupling limit $t ' = 0$ and {c)-d) the corresponding energy spectrum.
}
}
\label{fig:StrongCoupling}
\end{figure}

In the absence of $\Delta$, each state with even $N_{\rm tot}$ is degenerate, with allowed values $n_d = 0,2,4,\dots, 2k$ and all possibilities to distribute these electrons over the topological edge states. Similarly, the states with odd $N_{\rm tot}$ are also degenerate with $n_d = 1,3,5,\dots,2k-1$ allowed. The presence of $\Delta$ lifts the degeneracy as it allows to connect different states and favors a single BCS-like ground state~${\ket{\text{BCS}}_d}$ in the even sector (see supplement \cite{SuppMat} for details). In the odd sector, there are $2k$ ground states given in which one of the $k$ spin-degenerate  boundary states is singly occupied, while the remaining $k-1$ are occupied by a BCS-like state, see  Fig.~\ref{fig:StrongCoupling}, a). The ground state energy of the even sector is, 
\begin{equation} 
    E_{\rm even} (N_{\rm tot})  = E_C (N_{\rm tot} - N_g) ^2 - \Delta k, \label{eq:energies}
\end{equation}
while $E_{\rm odd} = E_{\rm even} + \Delta$. These energies are plotted in panel b) of Fig.~\ref{fig:SetupStates} (there, all energies $E$ are measured with respect to $-\Delta k$).

In the  $2k$-fold degenerate odd sector the quantum dot acts as an effective $Sp(2k)$ impurity. 
We will therefore consider $N_g$ close to 1, where the $2k$ odd parity states with $N_{\rm tot} =1 $ are lowest in energy 
while the lowest excited states (with $N_{\rm tot} = 0,2$) are separated by an energy gap
$\Delta E_\pm = E_{\rm even}({N_{\rm tot} =1\pm 1})-E_{\rm odd} ({N_{\rm tot} =1})$.
To derive the effective Kondo interaction, we next consider  tunneling between the electrons on the dot and the first site ($a = 0$) of the lead,  
$H_t = - \sum_{i = 1}^k\sum_{\sigma=\uparrow,\downarrow} t_{i} c^\dagger_{0,i, \sigma} d_{i \sigma} + \text{H.c.}$.   
At low temperatures and bias voltages in the weak tunneling limit, $k_B T, e V, t_i \ll \Delta E_{\pm}$, the dot occupation cannot change and $H_t$ induces an effective Kondo interaction in second order perturbation theory. When we fine-tune all $t_i = t \ ( \forall i = 1, \dots k)$, we get \cite{SuppMat},
\begin{equation}
H_{\rm eff} =
 -  \lambda_1  \qty(d^\dagger c_0)( c_0^\dagger d) -  \lambda_2 \qty( d^\dagger \sigma_y  c_0^* )\qty( c_0^T \sigma_y d ), \label{eq:effecH}
\end{equation} 
where $c^{*}_{a}=(c^\dagger_{a})^T$ and Gutzwiller projection to the $2k$ $N_{\rm tot} = 1$ states  is understood. The coupling constants are $\lambda_1=2 t^2/(\Delta E_-)>0$ and $\lambda_2=2 t^2/(\Delta E_+)>0$. Exactly at $N_g=1$, and after using the completeness relation of symplectic generators, $\sum_A T_A^{ij} T_A^{kl} = [\delta_{il} \delta_{jk} - (\sigma_y)_{ki} (\sigma_y)_{jl}]/2$,  Eq.~(\ref{eq:effecH}) becomes a Kondo-type interaction, Eq.~(\ref{eq:HKondo}), with (bare) coupling constant $\lambda=2\lambda_1=2\lambda_2=4t^2/(E_C-\Delta)$. As we will see below, the anisotropy of tunneling strength $t_i$ is irrelevant.

\PRLSection{Weak and strong coupling.} 
Perturbation theory in the Kondo term $H_K$ leads to the usual logarithmic divergence at second order~\cite{hewson1997kondo}. 
We therefore use the renormalization group (RG) technique to analyze Eq.~(\ref{eq:HKondo}) upon lowering the bare bandwidth/cutoff $D_0 \sim {E_C - \Delta} $ to a running cutoff $D = D_0 e^{-l}$~\cite{Anderson_1970,kogan2019poor}. We find the RG equation, 
\begin{equation}
\frac{{\rm d}\lambda}{{\rm d}l}=(k+1)\rho_0\lambda^2, \label{eq:isoRG}
\end{equation}
where $\rho_0=(\pi\hbar v_F)^{-1}$ denotes the lead  density of states per spin per length and $v_F$ is the Fermi velocity. 
Equation~(\ref{eq:isoRG}) implies that $\lambda$ flows towards stronger coupling upon reducing the energy cutoff (set by, e.g., the temperature). 
We estimate the strong coupling scale to be $T_{K} \sim  (E_C-\Delta) e^{-1/[\rho_0 \lambda(D_{0}) 2(k+1)]}$ in terms of the bare coupling. 

Given that the isotropic weak-coupling fixed point ($\lambda \! = \! 0$) is unstable, with RG flow towards strong coupling, we will next investigate the stability of the strong-coupling fixed point, where the local Kondo interaction~(\ref{eq:HKondo}) is the dominant term in the Hamiltonian, and we can treat kinetic energy $t' \sim 1/\rho_0$ of the leads perturbatively. On the bare level, this corresponds to the limit $t'  \ll t^2 /\Delta E_\pm$ of the mesoscopic device introduced above, see Fig.~\ref{fig:StrongCoupling}a). 

We start by finding the unperturbed ground state of Eq.~(\ref{eq:HKondo}), without kinetic terms. Similarly to Nozi\`eres'~\cite{Nozieres1974} description of the conventional $SU(2)$ Kondo problem, the strong coupling ground state is given by singlets formed by the impurity and the conduction electrons. We systematically derived the spectrum~\cite{SuppMat} of this problem using representation theory, and additionally explicitly constructed the singlet ground state wave functions for all $k$ and the excited states for $k = 2,3$
, see {Fig.~\ref{fig:StrongCoupling} c),d)}. We find that the $Sp(2k)$ Kondo Hamiltonian is overscreened, with $k$ degenerate ground states at $t' = 0$, e.g. for $k=2$ these are the  $Sp(2k)$ singlets,

\begin{subequations}
\begin{align}
    \ket{N=1}_\text{singlet}= & -i (d^\dagger \sigma_y c_0^*) \;\ket{0}_{c} \otimes \ket{\text{BCS}}_{d} 
    , \label{eq:N1}\\
    \ket{N=3}_\text{singlet}=&  (d^\dagger c_0) \;\ket{4}_c \otimes \ket{\text{BCS}}_d,\label{eq:N3}
\end{align}
\end{subequations}
where $\ket{4}_{c}$ is the state in which all electronic states on the first site of the lead are filled, while $\ket{0}_{c}$ is the empty state. For generic $k$, the degeneracy is a consequence of the symplectic symmetry associated to superconductivity\cite{AltlandZirnbauer1997}: Given that $\ket{N = 1}_{\rm singlet}$ is a singlet, the states $(c^\dagger_0 \sigma_y c^*_0)\ket{N = 1}_{\rm singlet}, \dots, (c^\dagger_0 \sigma_y c^*_0)^{k-1}\ket{N = 1}_{\rm singlet}$ transform trivially under $Sp(2k)$, as well~\footnote{Under $Sp(2k)$: $c_a \rightarrow U c_a$ with $U^T \sigma_y U = \sigma_y$}, see Fig.~\ref{fig:StrongCoupling} b). The above states, Eq.~(\ref{eq:N1})--(\ref{eq:N3}), are related by particle-hole symmetry (PHS). More generally, PHS implies an inherent $SU(2)$ symmetry in Nambu space for the symplectic Kondo Hamiltonian~\cite{FlintColeman2008} which can be made apparent by writing the $Sp(2k)$ currents as symmetric form, 
\begin{align}
J^A=\frac{1}{2}\left(\begin{array}{cc}
c_0^{\dagger} & c_0^{T}(i\sigma_{y})\end{array}\right)\left(\begin{array}{cc}
T_A & \mathbf{0}\\
\mathbf{0} & T_A
\end{array}\right)\left(\begin{array}{c}
c_0\\
(-i\sigma_{y})c^{*}_0
\end{array}\right), \label{eq:J}
\end{align}
which is  invariant under  $SU(2)$ rotations in particle-hole space. 
We used here  the property $T_A^{T}=-\sigma_{y} T_A\sigma_{y}$ of $Sp(2k)$ generators. After having established the $t' = 0$ ground states, we now incorporate the nearest-neighbor hopping $H_\text{NN}= -t^{\prime} \sum_{i=1}^k\sum_{\sigma=\uparrow,\downarrow} (c_{0,i,\sigma}^{\dagger}c_{1,i,\sigma}+c_{1,i,\sigma}^{\dagger}c_{0,i,\sigma})$ as a perturbation to study the stability of the strong coupling fixed point. $H_\text{NN}$ will couple the degenerate strong coupling ground states, Eqs.~(\ref{eq:N1})--(\ref{eq:N3}), in second order perturbation theory, while preserving the $SU(2)$ symmetry.  Inspired by the $SU(2)$ symmetry in the particle-hole space [see Eq. (\ref{eq:J})] and the $k$ singlets distributing in all odd-number particle sectors \cite{SuppMat}, we thus conjecture that the strong coupling Hamiltonian takes the form of channel-isotropic $k$ channel Kondo model, 
\begin{equation}
    H_{\text{s}}=
    \tilde{\lambda} \mathbf{S} \cdot \sum_{i = 1}^k \mathbf{s}_i, 
    \label{eq:scsp4H}
\end{equation}
where the impurity $SU(2)$ spin-$(k-1)/2$ operator $\mathbf{S}$ acts in the $k-$dimensional subspace (spanned by Eqs.~(\ref{eq:N1})--(\ref{eq:N3}) for $k = 2$),  $\mathbf{s}_i=f_{i}^{\dagger}(\boldsymbol{\sigma}/2)f_{i}$ and $f_{i}=(f_{i\uparrow},f_{i\downarrow})^T
\equiv (c_{1,i,\uparrow}^{\dagger},c_{1,i,\downarrow})^T$ with $i=1, \dots k$ labeling the effective channel of conduction electrons. Since $S=(k-1)/2 < k/2$, the MCK Hamiltonian~(\ref{eq:scsp4H}) is overscreened~\cite{Nozieres1980}. We have explicitly proven the conjecture for $k = 2$ ($k = 3$) by second-order perturbation theory (Schrieffer-Wolff transformation), for which virtual fluctuations into the 62 (381) excited states lead to $ \tilde{\lambda} = 24 t^{\prime2}/(5\lambda) $ ($\tilde \lambda =128 t^{\prime2}/(21\lambda)$), respectively~\cite{SuppMat}. In this context it is also worthwhile to point out a hidden (larger) $Sp(2k)$ symmetry in the $k-$channel $SU(2)$ Kondo effect~\cite{PhysRevB.45.7918}.

Since the weak-coupling limit of the overscreened multi-channel $SU(2)$ model is unstable~\cite{Nozieres1980,PhysRevB.75.045129}, the above map relating it to the strong-coupling limit of the symplectic Kondo model implies also the instability of the latter fixed point, see Fig.~\ref{fig:SetupStates}e. Together with the instability of the weak-coupling fixed point of the $Sp(2k)$ Kondo problem, see Eq.~(\ref{eq:isoRG}), these findings indicate a single stable fixed point between the two, i.e. at an intermediate coupling. Our conjecture of a single fixed point is supported by the low-temperature impurity entropy (below) which is found to have the same value, when approaching from the weak  [$Sp(2k)$] and strong  [$k$-channel $SU(2)$] coupling sides. Since $Sp(2)$ is isomorphic to $SU(2)$, our model provides an example of the level-rank duality~\cite{Kimura2021} relating the weak and strong coupling theories. 

\PRLSection{Observables: thermodynamics.}
Above, we argued that  near strong coupling, the model can be mapped to an overscreened $k$-channel spin-$(k-1)/2$ Kondo model, which has a stable intermediate coupling fixed point. 
We can use the impurity entropy \cite{affleck1995conformal} to characterize the effective residual ground state degeneracy $g_k$ of this fixed point. 
The ground state degeneracy associated to screening a spin $(k-1)/2$ with $k$ spin-1/2 channels is well-known, $ g_k=2\cos[\pi/(k+2)]$ \cite{AndreiDestri1984, Tsvelick1985, TsvelickWiegmann1985, affleck1995conformal}. 
This result agrees with the impurity entropy of the $Sp(2k)$ Kondo problem, calculated using
CFT~\cite{Kimura2021} and Bethe Ansatz~\cite{KoenigTsvelik2022}.

In particular, we note that the case $k=3$ has $g_3 = (1 + \sqrt{5})/2 = \varphi $, the Golden ratio, indicating an emergent Fibonacci anyon. Crucially, in our symplectic Kondo model this  Fibonacci anyon occurs even in the single-channel case (in the sense that our model, Eq.~(\ref{eq:HKondo}), is of level 1) and is therefore not subject to instability due to channel anisotropy, unlike previous examples in the 3-channel Kondo~\cite{PhysRevB.45.7918,PhysRevLett.128.146803} and 2-channel topological Kondo~\cite{https://doi.org/10.48550/arxiv.2207.10105} models. 

Despite this appearance of the same anyon-like ground-state degeneracies and an unstable strong coupling fixed point which is equivalent to the $k$-channel spin-$(k-1)/2$ $SU(2)$ Kondo model, we emphasize that in our model due to PHS, not all operators of the $SU(2)$ Kondo model are 
effective. For example, the symplectic susceptibility involves the excitation of states outside the low-energy manifold Eqs.~\eqref{eq:N1},\eqref{eq:N3}, leading to less singular behavior than for the $SU(2)_k$ susceptibility~\cite{SuppMat}. More generally, we expect that the  irrelevant operator of scaling dimension $1 + 2 / (2+k)$ is forbidden for the dual Kondo problem, Eq.~\eqref{eq:scsp4H}.
This implies Fermi-liquid like temperature and field dependence of thermodynamic quantities, consistent with results~\cite{Kimura2021, MitchellAffleck2021,LibermanSela2021, KoenigTsvelik2022} based on the weak coupling Hamiltonian, Eq.~\eqref{eq:HKondo}. 

We note that CFTs in which certain operators are symmetry disallowed are well known in the theory of (e.g. confinement-deconfinement) phase transitions in gauge theories and usually denoted by an asterisk~\cite{SchulerLauchli2016,Sachdev2018,Vojta2018}. In view of the relationship between deconfining gauge theories and overscreened Kondo impurities~\cite{KoenigColeman2021}, we borrow the notation employed for the latter phenomena and denote the boundary CFT describing the dual Kondo problem, Eq.~\eqref{eq:scsp4H}, as $SU(2)_k^*$.

\PRLSection{Observables: transport.} We propose to test the non-trivial nature of the symplectic Kondo effect in a charge transport experiment across the mesoscopic island. As we explicitly demonstrate~\cite{SuppMat} using the CFT method~\cite{affleck1995conformal,PhysRevLett.67.161,affleck1990current,affleck1991critical,affleck1991kondo,PhysRevB.48.7297,ludwig1994exact,PhysRevLett.67.3160,PhysRevB.58.3794}, the fixed point off-diagonal conductance, Eq.~(\ref{eq:G}), of $Sp(2k)$ Kondo model and the spin-$1/2$, $k$-channel $SU(2)$ charge Kondo model~\cite{PhysRevB.57.R5579,PhysRevB.65.195101} are identical up to normalization~\footnote{The natural normalization is to use the maximal conductance of a multiterminal junction, which in the $Sp(2k)$ case is  $G_{\text{max}}^{Sp(2k)}=4e^{2}/(h k)$. Due to spin degeneracy,  $G_{\text{max}}^{Sp(2k)}$ is twice as large as the maximum conductance in the $SO(k)$ topological Kondo effect and four times as large as in the $k$-channel $SU(2)$ charge Kondo effect (the additional factor 2 is due to the inelastic cotunneling  nature of the latter~\cite{PhysRevB.52.16676,PhysRevB.65.195101}). 
}.  
Nevertheless, we emphasize  that  our result is valid far from the charge degeneracy points, in the regime of elastic cotunneling  akin to spin Kondo effect~\cite{Pustilnik_2004}. At low temperatures, $T \ll T_K$, near the intermediate coupling fixed point, the off-diagonal $Sp(2k)$ charge conductance is 
\begin{equation}
    G_{i\neq j}(T)=\frac{4e^2}{h k }\sin^{2}\left(\frac{\pi}{k+2}\right) \left[1 +c_{ij} \left(\frac{T}{T_K}\right)^2
    \right], 
    \label{eq:G}
\end{equation} 
where the $T=0$ value is obtained in Ref.~\cite{SuppMat}. 
For $k=2$ we have exactly half of the maximum conductance, analogous to halving of the conductance in the spin 2-channel Kondo effect~\cite{OregDGG2002}  
and also similar to the conductance in the quarter-filling $SU(4)$ Kondo model~\cite{PhysRevLett.95.067204}. 
The finite-temperature correction with its dimensionless coefficient $c_{ij}$ and the Kondo temperature $T_K$ are determined from the microscopic physics, see below Eq.~(\ref{eq:isoRG}) for the latter. The temperature dependent transconductance (including larger temperature regimes) is plotted in Fig.~\ref{fig:SetupStates} c). The exponent in the finite-temperature correction to $G_{ij}(T=0)$ is determined by the scaling dimension $\Delta_{\text{LIO}}$ of the leading irrelevant operator. Importantly, in the 1-channel $Sp(2k)$ model the leading irrelevant operator~\cite{Kimura2021,MitchellAffleck2021,LibermanSela2021} is local density-density interaction with  $\Delta_{\text{LIO}} = 2$, giving a FL-like temperature dependence while it is non-FL like for $SU(2)_k$. As explained above, also for $SU(2)_k^*$ the operator responsible for non-FL power-laws is absent and we expect the exponent in Eq.~(\ref{eq:G}) to be the same regardless of whether we approach the stable intermediate ($T= 0$) fixed point from weak or strong coupling. The exotic zero-temperature conductance value $G_{i \neq j}(0)$ reminiscent of the multi-channel charge Kondo effect~\cite{PhysRevB.52.16676,PhysRevB.57.R5579,PhysRevB.65.195101,2015Natur.526..233I,2018Sci...360.1315I} together with FL corrections to it are unique signatures of the $Sp(2k)$ intermediate fixed point.

\PRLSection{Effect of anisotropy and PHS breaking. }  When deriving the $Sp(2k)$ Kondo interaction, we required fine tuning of the tunneling strengths $t_i=t$ ($\forall \, i=1,\cdots,k$) and particle-hole symmetry $N_g=1$. Although these parameters can be controlled in experiments, we will discuss next what happens when we deviate from the requirements. We  show that the former requirement can be relaxed but the deviation from PHS will drive the system towards an $SU(2k)$ Kondo fixed point. Let us first discuss the anisotropy of the tunneling amplitudes, while keeping the system PHS~\cite{SuppMat}. When we consider the anisotropic version of the effective Hamiltonian  Eq.~(\ref{eq:effecH}), the anisotropic tunneling strength $t_i >0$ can be  absorbed into the operators $\tilde{d}=\eta d$ and $\tilde{c}_0=\eta c_0$, where $\eta=\mathbb{I} \otimes \text{diag}(\sqrt{t_{1}},\cdots,\sqrt{t_{k}})/\sqrt{t}$, where $t$ now denotes the geometric mean of $t_i$. The anisotropic Hamiltonian then takes the same form as Eq.~(\ref{eq:effecH}), with the replacement $d,c_0\to \tilde{d}, \tilde{c}_0$. Upon using the completeness relation and restoring the physical operators $d,c_0$, we obtain transformed generators $\eta T_A \eta $ in the operators $S^A$ and $J^A$. The transformed generators are still $Sp(2k)$ generators because the matrix $\eta$ commutes with $(\sigma_{y}\otimes\mathbb{I})$, thus $(\sigma_{y}\otimes\mathbb{I})(\eta T_A \eta)^{T}(\sigma_{y}\otimes\mathbb{I})=-\eta T_A \eta$ according to the properties of $Sp(2k)$ generators \cite{footnote1}. Then, we can expand the transformed generators by the original generators: $\eta T_A \eta=\sum_{B}\kappa_{AB}T^{B}$. From this we see that the anisotropy of tunneling amplitudes is equivalent to the ``exchange'' anisotropy of the $Sp(2k)$ Kondo model, $H_K = \lambda \sum_{A,B}\kappa_{AB}S^A J^B$. Using the generalized version~\cite{SuppMat}, weak anisotropies $\vert \kappa_{AB} -\delta_{AB} \vert \ll 1$, can be shown to be irrelevant on general grounds. The same situation occurs with $SO(M)$, in the topological Kondo model~\cite{PhysRevLett.109.156803,PhysRevLett.110.216803, PhysRevB.94.235102,PhysRevResearch.2.043228,PhysRevB.97.235139}, where the isotropic direction dominates the RG flow. We note however that in the effective strong coupling multichannel $SU(2)$ model, Eq.~(\ref{eq:scsp4H}),   time-reversal symmetric tunneling anisotropy (unequal $t^{\prime}_i$) corresponds to channel anisotropy which is a relevant perturbation. Thus, the strong coupling multichannel Kondo physics requires fine-tuning of the $Sp(2k)$ symmetry. 

While at weak coupling anisotropy in the tunnel-couplings is harmless, the $Sp(2k)$ is more sensitive to breaking of PHS. We first consider $N_g \neq 1$ ($\lambda_1 \neq \lambda_2$) in Eq.~(\ref{eq:effecH}), whilst still requiring $t \ll \Delta E_{\pm}$ (regime of pink shading of Fig.~\ref{fig:SetupStates} b)). Then, we can rewrite Eq.~(\ref{eq:effecH}) as a potential scattering term for conduction electrons and an anisotropic $SU(2k)$ Kondo interaction. This $SU(2k)$ Kondo model {is exactly screened and} has a FL fixed point, and thus the non-FL fixed point of the $Sp(2k)$ Kondo model will be unstable. An example with $k=2$ has been discussed in Ref.~\cite{LibermanSela2021}. Also, the term %
arising from $\lambda_1 \neq \lambda_2$
maps to an effective magnetic field in the $SU(2)$ Kondo model in the strong coupling regime, similar to the case in charge Kondo~\cite{PhysRevB.52.16676}. {Near the intermediate coupling fixed point} such a perturbation is  relevant, with a scaling dimension $\Delta_H = 2 / (2+k)$, and drives the system to a FL fixed point~\cite{PhysRevB.45.7918}. 
Hence, we conclude that the PHS breaking anisotropy ($N_g \neq 1$) is relevant. 
As $N_g$ is further detuned from unity to a regime $t \sim \Delta E_\pm$ and further, we first enter an $SU(2k)$ mixed valence regime (dark gray in Fig.~\ref{fig:SetupStates} b), in which odd and even parity states are of comparable energy, and ultimately reach the regime in which the impurity ground state is non-degenerate. In the infrared, FL behavior persists, see Fig.~\ref{fig:SetupStates} d).  

\PRLSection{Summary and Conclusions.}
In summary, we proposed a mesoscopic implementation of the symplectic Kondo effect, in which the group  $Sp(2k)$ naturally describes spin-1/2 fermions in $k$ orbitals in a Coulomb blockaded island hosting $k$ spinful topological  Andreev states. We couple each Andreev state to a spinful fermion lead and found the symplectic Kondo Hamiltonian Eq.~\eqref{eq:HKondo} for an odd-parity charge state of the Coulomb blockaded island. 

Interesting open questions about the symplectic Cooper pair box setup include the Coulomb blockaded transport beyond $N_g = 1$ and complementary analytical, numerical and experimental studies which should help shed light on the anyonic signatures and their quantum-information theoretic potential.

\PRLSection{Acknowledgments.}
It is a pleasure to thank P. Coleman, Y. Komijani, M. Lotem, J. Schmalian, A. Schnyder, P. Simon, A.M. Tsvelik for useful discussions. 
This work was initiated at the Aspen Center for Physics, which is supported by National Science Foundation grant PHY-1607611. This material is based upon work supported by the U.S. Department of Energy, Office of Science, National Quantum Information Science Research Centers, Quantum Science Center.

\bibliography{apssamp}

\begin{thebibliography}{75}
\bibitem [{foo()}]{footnote1}%
  \BibitemOpen
  \href@noop {} {}\bibinfo {note} {The generators are $\mathbb{I} \otimes \mi
  \vec{A} $, $\sigma_x \otimes \vec{S} $, $\sigma_y \otimes \vec{S} $ and
  $\sigma_z \otimes \vec{S} $ where $\vec{A}$ are $k(k-1)/2$ real $k\times k$
  antisymmetric matrices, $\vec{S}$ are $k(k+1)/2$ real $k\times k$ symmetric
  matrices and $\sigma_{x,y,z}$ are Pauli matrices. They satisfy
  $(\sigma_{y}\otimes\mathbb{I})(T_A)^{T}(\sigma_{y}\otimes\mathbb{I})=-T_A$.
  We normalize the generators as $\Tr(T_A T_B) =\delta_{AB}$.}\BibitemShut
  {Stop}%
\end{thebibliography}

\clearpage
\newpage

\setcounter{equation}{0}
\setcounter{figure}{0}
\setcounter{section}{0}
\setcounter{table}{0}
\setcounter{page}{1}
\makeatletter
\renewcommand{\theequation}{S\arabic{equation}}
\renewcommand{\thepage}{S-\arabic{page}}
\renewcommand{\thesection}{S\arabic{section}}
\renewcommand{\thefigure}{S\arabic{figure}}
\renewcommand{\thetable}{S-\Roman{table}}

\begin{widetext}
\begin{center}
Supplementary materials on \\
\textbf{``Topological symplectic Kondo effect''}\\
Guangjie$^{1}$, Elio J. K\"onig$^{2}$, Jukka I. V{\"a}yrynen$^{1}$, \\ 
$^{1}$ \textit{Department of Physics and Astronomy, Purdue University, West Lafayette, Indiana 47907, USA}\\
$^{2}$ \textit{Max Planck Institute for Solid State Research, D-70569 Stuttgart, Germany}
\end{center}

These supplementary materials contain details about the implementation in Sec.~\ref{app:Implementation}, a discussion of symmetry breaking anisotropies in  Sec.~\ref{app:SymmetryAndAnisotropy}, the strong coupling solution in Sec.~\ref{app:StrongCoupling}, and the derivation of the conductance matrix near the intermediate and weak coupling fixed points in  Sec.~\ref{app:conductance}. 

\section{Derivation of the $Sp(2k)$ Kondo interaction}
\label{app:Implementation}
In this section, we summarize an implementation which is somewhat 
reminiscent of a topological Kondo effect based on Eq.~\eqref{eq:tunnelH} of the main text, reproduced here for convenience:
\begin{equation}
H_{\text{d}} = E_C (2 \hat{N}_C + \hat n_d - N_g)^2  
    - \frac{1}{2} \Delta \sum_{i = 1}^k \sum_{\sigma \sigma'} [\me^{-\mi \phi} d^\dagger_{i,\sigma} (\sigma_y)_{\sigma \sigma'} d^\dagger_{i, \sigma'} + \text{H.c.}]. \label{Seq:Hdot}
\end{equation}
As mentioned, this Hamiltonian preserves $2 \hat{N}_C + \hat n_d = \hat N_{\rm tot}$ and for $\Delta = 0$ each state by $N_{\rm tot}$ even is degenerate, with $n_d = 0,2,4,\dots, 2k$ and all potential possibilities to distribute these electrons over the topological edge states. Similarly the states with $N_{\rm tot}$ odd  are also degenerate with $n_d = 1,3,5,\dots,2k-1$. The presence of $\Delta$ lifts the degeneracy as it allows to connect different states. 

\subsection{The ground state of the tunneling Hamiltonian}

In the following we keep a closed subset of states with $N_{\rm tot}$ odd and even, of which we expect that it will contain the lowest energy state in that sector.
In the even parity sector, the closed set of states under consideration are generated by repeatedly applying $e^{- i \phi} (d^\dagger \sigma_y d^*)_i $ to the vacuum. Assuming $i_1 \neq i_2 \dots \neq i_{n_d/2}$ the general states are
\begin{equation}
\ket{n_d:i_1, i_2 \dots, i_{n_d/2}} = \frac{1}{2^{n_d/2}} e^{- i n_d \phi/2} (d^\dagger \sigma_y d^*)_{i_1}(d^\dagger \sigma_y d^*)_{i_2} \dots (d^\dagger \sigma_y d^*)_{i_{n_d/2}} \ket{0}.
\end{equation}
In total, there are $\mathcal N_k(n_d) = \left (\begin{array}{c} k \\ n_d/2  \end{array} \right)$ of such states at given even $n_d$ and a total of $\sum_{n_d = 0,2, \dots 2k}\mathcal N_k(n_d) = 2^k$ such states. For example the leading terms are
\begin{align}
\ket{0};\ \
\ket{2:i} = \frac{1}{2} e^{- i \phi} (d^\dagger \sigma_y d^*)_i \ket{0}; \ \
\ket{4:i_1,i_2}= \frac{1}{2^2} e^{- i 2\phi} (d^\dagger \sigma_y d^*)_{i_1}(d^\dagger \sigma_y d^*)_{i_2} \ket{0} .
\end{align}
In the odd parity sector, the closed set of states can be obtained by repeatedly acting $e^{- i \phi} (d^\dagger \sigma_y d^*)_i $ on the states with one electron being filled. Using $i_1 \neq i_2 \neq i_{[n_d-1]/2}$ we find
\begin{equation}
\ket{n_d: (i_0, \sigma), i_1, \dots i_{[n_d-1]/2}} = \frac{1}{2^{[n_d-1]/2}} e^{- i [n_d-1]\phi/2} d^\dagger_{i_0,\sigma} (d^\dagger \sigma_y d^*)_{i_1}(d^\dagger \sigma_y d^*)_{i_2} \dots (d^\dagger \sigma_y d^*)_{i_{[n_d-1]/2}} \ket{0}
\end{equation}
For a given $(i_0, \sigma)$, there are $\left (\begin{array}{c} k-1 \\ (n_d-1)/2 \end{array} \right)$ such states for each given $n_d$ and $2^{k-1}$ states in total.
For example the leading such terms are
\begin{align}
\ket{1: (i_0, \sigma)} =  d^\dagger_{i_0,\sigma} \ket{0}; \ \
\ket{3: (i_0, \sigma), i_1} = \frac{1}{2^{[n_d-1]/2}} e^{- i  \phi} d^\dagger_{i_0,\sigma} (d^\dagger \sigma_y d^*)_{i_1} \ket{0},
\end{align}
where $i_1 \neq i_0$ in the last line.

The set of parity states listed above is closed under the action of $H_{\text{d}}$, i.e. the states are mutually resonating into one-another with amplitude $-\Delta$. The choice of the above states is such that this quantum resonation can minimize the ``kinetic'' energy of hopping from one state to the other the most. In particular, consider an even parity state with $n_d$ boundary electrons. There are $n_d/2$ ways to reduce the number of boundary electrons by two. Moreover, there are $k-n_d/2$ ways to increase the number of boundary electrons by two. In total, the outlined set of even parity states forms a tight-binding Hamiltonian on a hypercube with $2^k$ corners. Each corner has coordination number $k$ and the quantum states are hybridized with amplitude $- \Delta$ on each bond. Similarly, for an odd parity state with $n_d+1$ boundary electrons in total and fixed $(i_0,\sigma)$. There are $n_d/2$ ways to reduce the number of boundary electrons by two. Moreover, there are $k-1-n_d/2$ ways to increase the number of boundary electrons by two. In total, the outlined set of odd parity states at given $(i_0,\sigma)$ forms a tight-binding Hamiltonian on a hypercube with $2^{k-1}$ corners each corner has coordination number $k-1$ and the quantum states are hybridized with amplitude $- \Delta$ on each bond.

Assuming that the groundstate of this tight binding Hamiltonian is the state with equal weight on all corners of the hypercube, 
\begin{align}
       \ket{N_{\rm tot}}_{\rm even} & = \frac{e^{i \frac{N_{\rm tot}}{2}\phi}}{\sqrt{2^k}} \prod_{i = 1}^k \left [ 1 + \frac{e^{- i\phi}(d^\dagger \sigma_y d^\dagger)_{i}}{2}\right] \ket{0} = \frac{e^{i \frac{N_{\rm tot}}{2}\phi}}{\sqrt{2^k}} e^{\sum_{i = 1}^k\frac{e^{- i\phi}(d^\dagger \sigma_y d^\dagger)_{i}}{2}} \ket{0}, \label{Seq:evenstate}\\
        \ket{N_{\rm tot}, (i_0,\sigma)}_{\rm odd} & = d^\dagger_{i_0,\sigma} \frac{e^{i (\frac{N_{\rm tot}-1}{2})\phi}}{\sqrt{2^{k-1}}} \prod_{i = 1}^k \left [ 1 + \delta_{i \neq i_0} \frac{e^{- i\phi}(d^\dagger \sigma_y d^\dagger)_{i}}{2}\right] \ket{0} = \frac{d^\dagger_{i_0,\sigma}  e^{i (\frac{N_{\rm tot}-1}{2})\phi}}{\sqrt{2^{k-1}}} e^{\sum_{i = 1}^k \delta_{i \neq i_0}\frac{e^{- i\phi}(d^\dagger \sigma_y d^\dagger)_{i}}{2}} \ket{0}, \label{Seq:oddstate}
\end{align}
we find
\begin{align}
   \ket{N_{\rm tot}}_{\rm even}:  E_{\rm min, even} (N_{\rm tot}) & = E_C (N_{\rm tot} - N_g) ^2 - \Delta k,  \\
    \ket{N_{\rm tot}, (i_0,\sigma)}_{\rm odd}:    E_{\rm min, odd} (N_{\rm tot}) & = E_C (N_{\rm tot} - N_g)^2 - \Delta (k-1).
\end{align}
In the main text, in particular in Eqs.~\eqref{eq:N1},~\eqref{eq:N3}, we use the notation $\ket{\text{BCS}}_d = \ket{N_{\rm tot}=0}_{\rm even}$. These are the ground states within the sector of a given $N_{\rm tot}$. In each such sector, there are states which are order $\Delta$ higher. Note that the many-body ground state in even parity sector are non-degenerate, while the ground state in the odd-parity sector has degeneracy $2k$. This concludes the derivation of the groundstates and Eq.~\eqref{eq:energies} of the main text.

\subsection{The superexchange calculations}

As mentioned in the main text, we consider $N_g$ closest to 1. Hopping onto/off the island connects the $2k$ ground states with the states $N_{\rm tot} = 0,2$ and energy $\Delta E_\pm$ higher,
\begin{align}
   \Delta E_+ 
   = E_C (3 - 2 N_g) - \Delta,\ \
      \Delta E_- 
      = E_C (2 N_g-1) - \Delta
\end{align}

The matrix elements between the $2k$ lowest energy states characterized by $(i_0, \sigma)$ and the states with even parity is
\begin{align}
&\bra{N_{\rm tot} = 1: (i_0, \sigma_0)} d^\dagger_{i,\sigma}  \ket{N_{\rm tot} = 0} = \sqrt{2} \delta_{ii_0} \delta_{\sigma \sigma_0},\ \
\bra{N_{\rm tot} = 1: (i_0, \sigma_0)} d_{i,\sigma}  \ket{N_{\rm tot} = 2} = - i \sigma \sqrt{2} \delta_{ii_0} \delta_{\bar{\sigma} \sigma_0}. \\
&\bra{N_{\rm tot} = 1: (i, \sigma_0)} H_t  \ket{N_{\rm tot} = 0} = - t_{i_0} \sqrt{2} c_{i_0 \sigma_0},\ \ \bra{N_{\rm tot} = 1: (i, \sigma_0)} H_t  \ket{N_{\rm tot} = 2} = i \sigma_0 t_{i_0} \sqrt{2} c_{i_0 \bar{\sigma}_0}^\dagger .
\end{align}

We thus find (we drop the label ``$N_{\rm tot}$'' from the quantum states)
\begin{align}
H_{\rm eff}  =& - \ket{1:(i,\sigma)}\bra{1:(i,\sigma)} H_t \ket{0} \frac{1}{\Delta E_-} \bra{0} H_t \ket{1: (i', \sigma')} \bra{1: (i', \sigma')} \notag\\
& - \ket{1:(i,\sigma)}\bra{1:(i,\sigma)} H_t \ket{2} \frac{1}{\Delta E_+} \bra{2} H_t \ket{1: (i', \sigma')} \bra{1: (i', \sigma')} \notag\\
= & - \sum_{ii'} \left [\frac{2t_i t_{i'}}{\Delta E_-} (d^\dagger c)_i(c^\dagger d)_{i'} +  \frac{2t_i t_{i'}}{\Delta E_+} (d^\dagger \sigma_yc^\dagger)_i(c \sigma_y d)  _{i'} \right] \label{eq:HeffAniso}
\end{align}
When we fine tune $N_g=1$ and all $t_i = t,\ \forall i = 1, \dots k$, we obtain Eq.~\eqref{eq:effecH} and ultimately Eq.~\eqref{eq:HKondo} of the main text using the completeness relation
\begin{equation}
    \sum_{A}(T_A)_{af}(T_A)_{be}=[\delta_{ae}\delta_{bf}-(\sigma_{y}\otimes\mathbb{I})_{ba}(\sigma_{y}\otimes\mathbb{I})_{fe}]/2. \label{eq:app:Completeness}
\end{equation}

\section{Symmetry-breaking anisotropies}
\label{app:SymmetryAndAnisotropy}

In this section we summarize symmetry breaking of the symplectic topological Kondo model, starting with tunneling anisotropy (Sec.~\ref{Ssec:tunneling_aniso}) and ending with particle-hole symmetry breaking (Sec.~\ref{Ssec:PHSBreaking}). 

\subsection{Anisotropic $Sp(2k)$ Kondo model from unequal tunneling amplitudes} \label{Ssec:tunneling_aniso}
We first consider the situation in which unequal hopping $t_i\neq t_j$ occurs for at least some $i \neq j$. The  tunnel-anisotopic  Hamiltonian, Eq.~\eqref{eq:effecH}, can be expressed as
\begin{align}
H_{{\rm eff}}= 
-\frac{2 t^2}{\Delta E_{-}}\sum_{ii'}(\tilde{d}_{i}^{\dagger}\tilde{c}_{i})(\tilde{c}_{i'}^{\dagger}\tilde{d}_{i'})-\frac{2 t^2}{\Delta E_{+}}\sum_{ii'}(\tilde{d}_{i}^{\dagger}\sigma_{y}\tilde{c}_{i}^{*})(\tilde{c}_{i'}^{T}\sigma_{y}\tilde{d}_{i'}),
\end{align}
where $d_{i}=(d_{i,\uparrow},d_{i,\downarrow})^{T}$, $\tilde{d}_{i}=\sqrt{t_{i}}d_{i}/\sqrt{t}$, and $t = (\prod_i t_i)^{1/k} $. 
Let us first consider $N_{g}=1$ which makes $\Delta E_{-}=\Delta E_{+}$,
then
\begin{align}
H_{{\rm eff}}= -\frac{2t^2}{\Delta E_{-}}\sum_{ii'}[(\tilde{d}_{i}^{\dagger}\tilde{c}_{i})(\tilde{c}_{i'}^{\dagger}\tilde{d}_{i'})+(\tilde{d}_{i}^{\dagger}\sigma_{y}\tilde{c}_{i}^{*})(\tilde{c}_{i'}^{T}\sigma_{y}\tilde{d}_{i'})] 
=\frac{4t^2}{\Delta E_{-}}\sum_{A}\sum_{a,b,e,f}\tilde{d}_{a}^{\dagger}(T_A)_{af}\tilde{d}_{f}\tilde{c}_{b}^{\dagger}(T_A)_{be}\tilde{c}_{e}.
\end{align}
We again used the completeness relation Eq.~\eqref{eq:app:Completeness}.
Here, $\tilde{d}_{a}$ is the $a$th component of $(\tilde{d}_{1,\cdots,k;\uparrow},\tilde{d}_{1,\cdots,k;\downarrow})^{T}.$
If we define a matrix $\eta=
\text{diag}(\sqrt{t_{1}},\cdots,\sqrt{t_{k}},\sqrt{t_{1}},\cdots,\sqrt{t_{k}})/\sqrt{t}$,
we will have
\begin{align}
H_{{\rm eff}}
=  \frac{4t^2}{\Delta E_{-}}\sum_{A}\sum_{a',b',e',f'}\sum_{a,b,e,f}d_{a'}^{\dagger}(\eta T_A\eta)_{a'f'}d_{f'}c_{b'}^{\dagger}(\eta T_A \eta)_{b'e'}c_{e'}.
\end{align}
The transformed generators $\eta T_A \eta$ are still $Sp(2k)$
generators since $(\sigma_{y}\otimes\mathbb{I})(\eta T_A \eta)^{T}(\sigma_{y}\otimes\mathbb{I})=\eta(\sigma_{y}\otimes\mathbb{I})(T_A)^{T}(\sigma_{y}\otimes\mathbb{I})\eta=-\eta T_A \eta$
according to the definition $(\sigma_{y}\otimes\mathbb{I})(T_A)^{T}(\sigma_{y}\otimes\mathbb{I})=-T_A$.
Thus, $\eta T_A\eta=\sum_{B}\kappa_{AB} T^{B}$ (or equivalently $\kappa_{AB} = \tr[T_A \eta T_B \eta] = \kappa_{BA}$), which gives an anisotropic $Sp(2k)$ Kondo model with a Hamiltonian 
\begin{equation}
    H_K = \sum_{A,B} \lambda_{AB} S^A J^B, \quad  \lambda_{AB} =  \lambda \, \kappa_{AB}. \label{seq:HKaniso}
\end{equation}
We now briefly review that the weak coupling RG equations for $\lambda_{AB}$ generically lead to emergent symmetry $\lambda_{A B} = \delta_{AB} \lambda$. 
For generality, we do this for general group with structure constants $f^{ABC}$. 
The second-order $\beta$-function gives the RG equation~\cite{kogan2019poor} 
\begin{equation}
\frac{\mathrm{d}\lambda_{AB}}{\mathrm{d}l}=\sum_{CDEF} f^{ACE} f^{BDF} \lambda_{CD}\lambda_{EF} 
\end{equation}
with $\lambda_{AB}$ a symmetric matrix, see above Eq.~\eqref{seq:HKaniso}. 
We rotate into its eigenbasis (denoted by lower-case $a,b,c = 1, \dots D$, where $D$ is the group dimension), so that the RG equations become
\begin{equation}
\frac{d\lambda_{a}}{dl} = \sum_{bc}f_{abc}^2\lambda_{b} \lambda_{c}.
\end{equation}
We want to know the beta function near some point $\lambda_a = \lambda \ \forall a$ (it is not necessarily an RG fixed point, but rather an arbitrary point on the isotropy line in the space of coupling constants). Then, using $\lambda_a = \lambda + \delta \lambda_a$ we get to leading order
\begin{equation}
\frac{d \vec{\delta \lambda}}{dl} 
=  \lambda \underline N [\sqrt{D} \lambda \hat e_1 + 2\vec{\delta \lambda}].
\end{equation}
Here, we grouped all $\delta \lambda_a$ into a vector, and introduced $\hat{e}_1 = (1,1,\dots, 1)/\sqrt{D}$. The matrix $N_{ab} = \sum_c f_{abc}^2$ has all diagonal elements zero (it is thus traceless), contains only positive entries and has the property that $\hat e_1$ is an eigenvector:  
\begin{equation}
N \hat e_1 = \mu_1 \hat e_1, \mu_1 = \sum_{abc}f_{abc}^2/D. 
\end{equation}
By the Perron-Frobenius theorem, we know that the unique largest eigenvalue of $N_{ab}$ has an eigenvector which can be chosen to have only positive entries. In the present case, we know that $\hat e_1$ is an eigenvector that fulfills this property and that all other eigenvectors have to be orthogonal to $\hat e_1$, and therefore necessarily contain negative entries. Thus, we conclude that $\mu_1$ is the largest eigenvalue of $N_{ab}$.

Next, we expand $\vec{\delta \lambda} = \sqrt{D} \lambda d_\alpha \hat e_\alpha$ where $\lbrace \hat e_\alpha \rbrace_{\alpha = 1}^D$ is the eigenbasis of $N_{ab}$. Note that, by assumption $d = \vert \vec d \vert \ll 1/\sqrt{D}$.

Then
\begin{equation}
\dot{d}_\alpha = \lambda [\mu_1 \delta_{\alpha,1} + 2\mu_\alpha d_\alpha].
\end{equation}
The RG equation for $\vec d = d \hat d$ (with $\hat d$ the unit vector) becomes 
\begin{equation}
\dot{d} \hat d+ d \dot{\hat d} = \lambda \left[\left (\begin{array}{c}
\mu_1 \\ 
0 \\ 
\vdots
\end{array} \right) + 2 \underline{\mu} d \hat d \right],
\end{equation} 
with $\underline \mu = \text{diag}(\mu_1, \dots, \mu_D)$,
or, equivalently, 
\begin{align}
\dot{d} &=  \lambda \hat d^T\left[\left (\begin{array}{c}
\mu_1 \\ 
0 \\ 
\vdots
\end{array} \right) + 2 \underline{\mu} d \hat d \right], \\
\dot{\hat d} &= \frac{\lambda}{d} \left [\mathbf 1 - \hat d \hat d^T	 \right ] \left[\left (\begin{array}{c}
\mu_1 \\ 
0 \\ 
\vdots
\end{array} \right) + 2 \underline{\mu} d \hat d \right]. \label{eq:hatddot}
\end{align}
As we primarily want to know the direction of flow near isotropy, we now concentrate on and expand the unit vector, 
\begin{align}
\hat d = \left (\begin{array}{c}
\sqrt{1 -\sum_{\alpha = 2}^D \delta d_\alpha^2} \\ 
\delta d_2 \\ 
\vdots \\ 
\delta d_D
\end{array} \right) &\approx  \left (\begin{array}{c}
1 \\ 
\delta d_2 \\ 
\vdots \\ 
\delta d_D
\end{array} \right),\\
 \left [\mathbf 1 - \hat d \hat d^T	 \right ] & \approx \left (\begin{array}{cc}
0 & - \vec{\delta d}^T \\ 
- \vec{\delta d} & \mathbf{1}
\end{array} \right),\\
\left[\left (\begin{array}{c}
\mu_1 \\ 
0 \\ 
\vdots
\end{array} \right) + 2 \underline{\mu} d \hat d \right] & \approx  \left (\begin{array}{c}
\mu_1 + 2d \mu_1 \\ 
2\mu_2 d \delta d_2 \\ 
\vdots
\end{array} \right),
\end{align}
so that the last $D-1$ equations in Eq.~\eqref{eq:hatddot} become
\begin{equation}
\dot{\vec{\delta d}} = - (\lambda/d)  \mu_1 (1 + 2d) \vec{\delta d} + 2 \lambda \underline \mu \vec{\delta d}
\end{equation}
or equivalently
\begin{equation}
\dot{\delta d_\alpha} = - [\mu_1/d +2(\mu_1-\mu_\alpha)] \lambda \delta d_\alpha, \quad \alpha = 2,\dots, D.
\end{equation} 
Since the square bracket is strictly positive (even once $d$ has increased to order unity), all deviations from the isotropic line are irrelevant.

\subsection{Particle-hole symmetry breaking anisotropy}\label{Ssec:PHSBreaking}
Since the tunneling strength anisotropy is irrelevant,  we can consider the isotropic case for tunneling strength but with $N_{g} \neq 1$, $\Delta E_{+}=E_{C}(3-2N_{g})-\Delta$
and $\Delta E_{-}=E_{C}(2N_{g}-1)-\Delta$. We define $\lambda_1=2t^2/\Delta E_{-}>0$ and $\lambda_2=2t^2/\Delta E_{+}>0$ and rewrite the effective Hamiltonian.
This anisotropy breaks the particle-hole symmetry.
\begin{align}
H_{{\rm eff}}= 
-\lambda_1 (d^{\dagger}c)(c^{\dagger}d) -\lambda_2 (d^{\dagger}h)(h^{\dagger}d) 
\end{align}
with $h_{m'}^{\dagger}=\sum_{m}(\mathrm{i}\sigma_{y}\otimes\mathbb{I})_{m'm}c_{m}$. By using the following identities, 
\begin{align}
&(d^{\dagger}c)(c^{\dagger}d)
=	\frac{1}{2k}\sum_{ab}d_{a}^{\dagger}d_{a}c_{b}c_{b}^{\dagger}-2\sum_{A}\sum_{aba'b'}d_{a}^{\dagger}\tau^{ab'}_Ad_{b'}\tau^{a'b}_Ac_{a'}^{\dagger}c_{b} \\
&(d^{\dagger}h)(h^{\dagger}d)
=	\frac{1}{2k}\sum_{ab}d_{a}^{\dagger}d_{a}c_{b}^{\dagger}c_{b}+2\sum_{A}\sum_{aba'b'}d_{a}^{\dagger}\tau^{ab'}_Ad_{b'}c_{b}^{\dagger}[(\sigma_{y}\otimes\mathbb{I})\tau_A(\sigma_{y}\otimes\mathbb{I})]_{ba'}^{T}c_{a'},
\end{align}
and the completeness relation of $SU(2k)$: $\sum_{A}\tau^{ab'}_A\tau^{a'b}_A=\frac{1}{2}\Big(\delta_{ab}\delta_{a'b'}-\frac{1}{2k}\delta_{ab'}\delta_{a'b}\Big)$, we get
\begin{align}
H_{{\rm eff}}=-\frac{\lambda_{1}}{2k}\sum_{ab}d_{a}^{\dagger}d_{a}c_{b}c_{b}^{\dagger}-\frac{\lambda_{2}}{2k}\sum_{ab}d_{a}^{\dagger}d_{a}c_{b}^{\dagger}c_{b}+
 \sum_{A}\sum_{aba'b'}d_{a}^{\dagger}\tau^{ab'}_Ad_{b'}c_{a'}^{\dagger}[2\lambda_{1}\tau_A-2\lambda_{2}(\sigma_{y}\otimes\mathbb{I})(\tau_A)^{T}(\sigma_{y}\otimes\mathbb{I})]_{a'b}c_{b} \label{eq:anisoHeff}
\end{align}
Since we have $\sum_{a}d_{a}^{\dagger}d_{a} = k$ when projected to the odd state, Eq.~(\ref{Seq:oddstate}), the first two terms of Eq.~(\ref{eq:anisoHeff}) are potential scattering terms for conduction electrons (they couple to the total density $\sum_{b}c_{b}^{\dagger}c_{b}$) and are present when $\lambda_1 \neq \lambda_2$.  The last term can be considered as an anisotropic $SU(2k)$ Kondo interaction by defining 
\begin{align}
 & \sum_{B}\kappa_{AB}\tau_{B}\equiv 2\lambda_{1}\tau_A-2\lambda_{2}(\sigma_{y}\otimes\mathbb{I})(\tau_A)^{T}(\sigma_{y}\otimes\mathbb{I})\\
\Rightarrow & \kappa_{AB}=2\Tr{[2\lambda_{1}\tau_A-2\lambda_{2}(\sigma_{y}\otimes\mathbb{I})(\tau_A)^{T}(\sigma_{y}\otimes\mathbb{I})]\tau_{B}},\ \Tr(\tau_A\tau_{B})=\frac{1}{2}\delta_{AB}.
\end{align}
By applying the second order RG equations of $SU(2k)$, we get
\begin{equation}
\frac{\mathrm{d}\lambda_{1}}{\mathrm{d}l}=2k\rho_{0}\lambda_{1}^{2}+2\rho_{0}\lambda_{1}\lambda_{2};\ \frac{\mathrm{d}\lambda_{2}}{\mathrm{d}l}=2k\rho_{0}\lambda_{2}^{2}+2\rho_{0}\lambda_{1}\lambda_{2}\Rightarrow\frac{\mathrm{d}(\lambda_{1}-\lambda_{2})}{\mathrm{d}l}=2k\rho_{0}(\lambda_{1}-\lambda_{2})(\lambda_{1}+\lambda_{2}).
\end{equation}
According to our definitions $\lambda_{1},\lambda_{2}>0$, we conclude that the anisotropy $N_{g}\neq1$ is relevant but less relevant than the isotropic direction. In the isotropic case, $H_{\text{eff}}$ becomes isotropic $Sp(2k)$ Kondo interaction $\lambda S^A J^A$ where $\lambda = 2\lambda_1=2\lambda_2$ and the isotropic RG equation is
\begin{equation}
 \frac{\mathrm{d}\lambda}{\mathrm{d}l} = (k+1) \rho_0 \lambda^2.
\end{equation}

At the strong coupling regime, we calculate the matrix element of $(d^{\dagger}c)(c^{\dagger}d)-(d^{\dagger}h)(h^{\dagger}d)$ for $Sp(2k)$ by using the unnormalized strong coupling singlet states Eq.~(\ref{eq:singletstats}):
\begin{equation}
\frac{\bra{N=2j-1}(d^{\dagger}c)(c^{\dagger}d)\ket{N=2j-1}}{\bra{N=2j-1}\ket{N=2j-1}}=2j,\ \frac{\bra{N=2j-1}(d^{\dagger}h)(c^{\dagger}h)\ket{N=2j-1}}{\bra{N=2j-1}\ket{N=2j-1}}=2k+2-2j.
\end{equation}
The matrix element for $(d^{\dagger}c)(c^{\dagger}d)-(d^{\dagger}h)(h^{\dagger}d)$ is $4S_{z}$ where $S_{z}$ is the $z$-component of spin $(k-1)/2$ generators: $\frac{1}{2}\text{diag}(k-1,k-3,\cdots,-k+3,-k+1)$. Hence, we conclude that $N_g \neq 1$ induces an extra magnetic term $S_z$ at the strong coupling regime. Both of the potential terms of $c$-fermions and the $SU(2k)$ Kondo term in Eq.~(\ref{eq:anisoHeff}) contribute to $2S_z$.

\section{Strong coupling Hamiltonian for symplectic Kondo model }
\label{app:StrongCoupling}
In this section we derive Eq.~\eqref{eq:scsp4H}, valid in the limit when the Kondo interaction $\lambda$ exceeds the bandwidth of the conduction electrons.

The $Sp(2k)$ Kondo Hamiltonian is  $H/
\lambda=\sum_{A}S_A J_A$ where
$S_A=\sum_{ij}d_{i}^{\dagger}T^{ij}_A d_{j}$ and $J_A=\sum^{kl}c_{k}^{\dagger}T^{kl}_A c_{l}$.
By using the Fierz identities \eqref{eq:app:Completeness} 
\begin{equation}
H/\lambda
=\frac{1}{2}\sum_{ijkl}[\delta_{il}\delta_{jk}-(\sigma_{y})_{ki}(\sigma_{y})_{jl}]d_{i}^{\dagger}d_{j}c_{k}^{\dagger}c_{l}
\end{equation}
The general states take the form $|m_{1}\cdots m_{K},p\rangle=c_{m_{1}}^{\dagger}\cdots c_{m_{K}}^{\dagger}d_{p}^{\dagger}|0\rangle$
where $m$s and $p$ can take values $1,2,\cdots,2k$.  In this section, the ``vacuum'' state $|0\rangle$ is understood to be the empty site of the lead and the  BCS state $\ket{\text{BCS}}_d = \ket{N_{\rm tot}=0}_{\rm even}$ for the dot. 

\subsection{Effective Hamiltonian of $Sp(4)$ Kondo model at strong coupling}

\subsubsection{Representations of current operator}

In the 1-particle sector, the matrix element of conduction electrons $J_A=\sum_{n_{1}n_{2}}c_{n_{1}}^{\dagger}(T_A)_{n_{1}n_{2}}c_{n_{2}}$ is simply $T_A$. Let us consider the 2-particle representation of it for $Sp(4)$ Kondo. We define the 2-particle states as $|\alpha\rangle=\sum_{m_{1}m_{2}}(A^{\alpha})_{m_{1}m_{2}}c_{m_{1}}^{\dagger}c_{m_{2}}^{\dagger}|0\rangle$. It is required that $A^{\alpha T}=-A^{\alpha}$ and $A^{\alpha\dagger}=A^{\alpha}$. The normalization condition is $\langle\beta|\alpha\rangle=2\Tr(A^{\beta}A^{\alpha})=\delta_{\beta\alpha}$
Thus, the 2-particle representation of generators are
\begin{equation}
J^{(2)}_A=\langle\beta|J_A|\alpha\rangle=\sum_{n_{1}n_{2}}\sum_{m_{1}m_{2}m_{1}'m_{2}'}(A^{\beta})_{m_{1}'m_{2}'}^{*}(T_A)_{n_{1}n_{2}}(A^{\alpha})_{m_{1}m_{2}}\langle0|c_{m_{2}'}c_{m_{1}'}c_{n_{1}}^{\dagger}c_{n_{2}}c_{m_{1}}^{\dagger}c_{m_{2}}^{\dagger}|0\rangle = 4\Tr(A^{\beta}T_AA^{\alpha}).
\end{equation}
Let us consider the following $A$ tensor
\begin{align}
[A^{\alpha=(\alpha_{1},\alpha_{2})}]_{m_{1}m_{2}} & =-\frac{\mathrm{i}}{2!}\delta_{m_{1}}^{\alpha_{1}}\delta_{m_{2}}^{\alpha_{2}}+\frac{\mathrm{i}}{2!}\delta_{m_{1}}^{\alpha_{2}}\delta_{m_{2}}^{\alpha_{1}}\ \text{with}\ \alpha_{1}<\alpha_{2}.
\end{align}
Here, $\alpha_1,\alpha_2,m_1,m_2=1,2,3,4$. 

\subsubsection{Strong coupling solution using representation theory}

We complete the square to compute the Kondo energy for $N$-particle sector 
\begin{equation}
2 H_K/\lambda=2\sum_{A}T_AJ^{(N)}_A=\sum_{A}[(T_A+J^{(N)}_A)^{2}-(T_A)^{2}-(J^{(N)}_A)^{2}],
\end{equation}
where $J^{(N)}$ is the current evaluated in the $N$ particle sector and we have replaced the matrix element of impurity operator $S^A$ by its 1-particle representation which is the fundamental representation $\mu_1$ associated to matrices $T_A$. While the trivial representation is denoted by $0$, higher dimensional representations are denoted $\mu_2, 2\mu_1$, where the five-dimensional representation $\mu_2$ is related to isomorphism with $SO(5)$. These $\mu_i$ are the highest weights of $Sp(4)$ Lie algebra.
Given that each representation has a well-defined, distinct Casimir invariant, we can readily calculate the eigenvalues and degeneracies of 
$H_K/\lambda$ and show the result in Tab.~\ref{tab:Sp4}
\begin{table}[t]
\begin{centering}
\begin{tabular}{|l|c|c|c|c|c|}
\hline 
$N$ &$\mu_{1}\otimes\mu_{1}\ (4\times4)$ & $\sum_{A}(T_A+J^{(1)}_A)^{2}$ & $-\sum_{A}(T_A)^{2}$ & $-\sum_{A}(J^{(1)}_A)^{2}$ & $ 2 H_K/\lambda$ \tabularnewline
\hline 
1,3 & $2\mu_{1}\ (10)$ & $6$ & $-5/2$ & $-5/2$ & $1$\tabularnewline
\hline 
1,3 & $\mu_{2}\ (5)$ & $4$ & $-5/2$ & $-5/2$ & $-1$\tabularnewline
\hline 
1,3 & $0\ (1)$ & $0$ & $-5/2$ & $-5/2$ & $-5$\tabularnewline
\hline 
\hline 
$N$ & $\mu_{1}\otimes(\mu_{2}\oplus0)\ [4\times(5+1)]$ & $\sum_{A}(T_A+J^{(2)}_A)^{2}$ & $-\sum_{A}(T_A)^{2}$ & $-\sum_{A}(J^{(2)}_A)^{2}$ & $2 H_K/\lambda$ \tabularnewline
\hline 
2 & $\mu_{1}+\mu_{2}\ (16)$ & $15/2$ & $-5/2$ & $-4$ & $1$\tabularnewline
\hline 
2 & $\mu_{1}\ (4)$ & $5/2$ & $-5/2$ & $-4$ & $-4$\tabularnewline
\hline 
2 & $\mu_{1}\ (4)$ & $5/2$ & $-5/2$ & $0$ & $0$ \tabularnewline
\hline 
\end{tabular}
\par\end{centering}
\caption{Decomposition of $Sp(4)$ Kondo interaction with dimensions and energies in the sectors of 1-3 lead electrons ($N$). The numbers in the brackets next to the representations are dimensions. The second $\mu_{1}$ (with $\sum_{A}(J^{(2)}_A)^{2}=0$) of 2-particle decomposition is from $\mu_{1}\otimes0$. The three particle decomposition follows from $N = 1$ by particle-hole conjugation.}
\label{tab:Sp4}
\end{table}
These results are the basis to Fig.~\ref{fig:StrongCoupling} c). we now discuss the wave functions associated to these states and thereby illustrate the meaning of different representations. 

\subsubsection{Eigenstates}

The lowest energy of all particle numbers comes from the singlet states
in 1 and 3-particle due to particle-hole symmetry. Now, let us define
the states in different particle numbers with impurity. The 0-particle
sector has states $|N=0,p\rangle=d_{p}^{\dagger}|0\rangle$. The 1-particle
states in natural basis are $|m,p\rangle=c_{m}^{\dagger}d_{p}^{\dagger}|0\rangle$
with $m,p=1\ \text{to}\ 4$. From the table above, the states are
16 dimensional, they are
\begin{equation}
|N=1,2\mu_{1},a\rangle=\sum_{mp}(s_A)_{mp}|m,p\rangle;\ |N=1,\mu_{2},b\rangle=\sum_{mp}(t^{b})_{mp}|m,p\rangle;\ |N=1,0\rangle=-\frac{\mathrm{\mathrm{i}}}{2}\sum_{mp}(\sigma_{y}\otimes\mathbb{I})_{mp}|m,p\rangle.
\end{equation}
Here $s_A(a=1,\cdots,21)$ are symmetric matrices and $t^{b}(b=1,\cdots,14)$
are antisymmetric matrices. Similarly, the 2-particle states are generally
defined as
\begin{equation}
|N=2,c\rangle=\sum_{\alpha p}(r^{c})_{\alpha p}|\alpha,p\rangle=\sum_{\alpha p}(r^{c})_{\alpha p}\sum_{m_{1}m_{2}}(A^{\alpha})_{m_{1}m_{2}}c_{m_{1}}^{\dagger}c_{m_{2}}^{\dagger}d_{p}^{\dagger}|0\rangle.
\end{equation}
We can define the three different decompositions above with different
sets of $c$.
\begin{align}
& (2 H_K/\lambda) |N=2,\mu_{1}+\mu_{2},c=1,\cdots,16\rangle=  |N=2,c=1,\cdots,16\rangle;\\
& (2 H_K/\lambda)|N=2,\mu_{1},c=17,\cdots,20\rangle=  -4 |N=2,c=17,\cdots,20\rangle;\\
& (2 H_K/\lambda) |N=2,\mu_{1},c=21,\cdots,24\rangle=  0.
\end{align}
The 3- and 4-particle sectors are particle-hole symmetric with the 1- and
0-particle sectors. The hole creation operator is defined as $h_{m'}^{\dagger}=\sum_{m}(\mathrm{i}\sigma_{y}\otimes\mathbb{I})_{m'm}c_{m}$.
The 4-particle sector has states $|N=4,p\rangle=d_{p}^{\dagger}|4,0\rangle$.
The 3-particle sector is the same as 1-particle sector except that
all the particle operators should be replaced by hole operators. For
example, the lowest energy states in this sector is
\begin{equation}
|N=3,0\rangle=-\frac{\mathrm{i}}{2}\sum_{m'p}(\sigma_{y}\otimes\mathbb{I})_{m'p}h_{m'}^{\dagger}d_{p}^{\dagger}|4,0\rangle=\frac{1}{2}\sum_{mm'p}(\sigma_{y}\otimes\mathbb{I})_{m'p}(\sigma_{y}\otimes\mathbb{I})_{m'm}c_{m}d_{p}^{\dagger}|4,0\rangle=-\frac{1}{2}\sum_{m}c_{m}d_{m}^{\dagger}|4,0\rangle.
\end{equation}

\subsubsection{Perturbative inclusion of the leads}

Now, we can calculate the effective Hamiltonian. The hopping energy
between the first and second site is $H_{12}=-t' \sum_{r}c_{1r}^{\dagger}c_{2r}+c_{2r}^{\dagger}c_{1r}$.
\begin{subequations}
\begin{align}
(H_{\text{eff}})_{11}= & \sum_{p'=1}^{4}\frac{|\langle N=0,p'|H_{12}|N=1,0\rangle|^{2}}{(-5\lambda/2-0)}+\sum_{c=1}^{16}\frac{|\langle N=2,c|H_{12}|N=1,0\rangle|^{2}}{(-5\lambda/2-\lambda/2)}+\sum_{c=17}^{20}\frac{|\langle N=2,c|H_{12}|N=1,0\rangle|^{2}}{(-5\lambda/2+4\lambda/2)} \notag\\
 & +\sum_{c=21}^{24}\frac{|\langle N=2,c|H_{12}|N=1,0\rangle|^{2}}{(-5\lambda/2-0)}.\\
(H_{\text{eff}})_{22}= & \sum_{p'=1}^{4}\frac{|\langle N=4,p'|H_{12}|N=3,0\rangle|^{2}}{(-5\lambda/2-0)}+\sum_{c'=1}^{16}\frac{|\langle N=2,c'|H_{12}|N=3,0\rangle|^{2}}{(-5\lambda/2-\lambda/2)}+\sum_{c'=17}^{20}\frac{|\langle N=2,c'|H_{12}|N=3,0\rangle|^{2}}{(-5\lambda/2+4\lambda/2)} \notag\\
 & +\sum_{c'=21}^{24}\frac{|\langle N=2,c'|H_{12}|N=3,0\rangle|^{2}}{(-5\lambda/2-0)}.\\
(H_{\text{eff}})_{12}= & [(H_{\text{eff}})_{21}]^{\dagger}= \sum_{c=1}^{16}\frac{\langle N=1,0|H_{12}|N=2,c\rangle\langle N=2,c|H_{12}|N=3,0\rangle}{(-5\lambda/2-\lambda/2)}\notag \\
 & +\sum_{c=17}^{20}\frac{\langle N=1,0|H_{12}|N=2,c\rangle\langle N=2,c|H_{12}|N=3,0\rangle}{(-5\lambda/2+4\lambda/2)}\notag\\
 & +\sum_{c=21}^{24}\frac{\langle N=1,0|H_{12}|N=2,c\rangle\langle N=2,c|H_{12}|N=3,0\rangle}{(-5\lambda/2-0)}.
\end{align}
\end{subequations}
We need to calculate the following terms
\begin{align}
&\langle N=0,p'|H_{12}|N=1,0\rangle=-\frac{\mathrm{i}t}{2}\sum_{r}(\sigma_{y}\otimes\mathbb{I})_{rp'}c_{2r}^{\dagger},\ \langle N=1,0|H_{12}|N=2,c\rangle =  -\mathrm{i}t\sum_{r}\sum_{\alpha p}(r^{c})_{\alpha p}[A^{\alpha}(\sigma_{y}\otimes\mathbb{I})]_{rp}c_{2r}^{\dagger},\\
&\langle N=2,c'|H_{12}|N=3,0\rangle=t\sum_{r}\sum_{\alpha p}(r'^{c'})_{\alpha p}^{*}(C^{\alpha})_{rp}^{*}c_{2r}^{\dagger},\ \langle N=3,0|H_{12}|N=4,p'\rangle=-\frac{t}{2}c_{2p'}^{\dagger},
\end{align}
where we defined $(C^{\alpha})_{rp'}=\frac{1}{2}\sum_{m_{1}'m_{2}'}\epsilon_{m_{2}'m_{1}'rp'}(A^{\alpha})_{m_{1}'m_{2}'}$.
Finally, we have
\begin{align}
& (H_{\text{eff}})_{11}=-\frac{t^{2}}{10\lambda}\sum_{r}c_{2r}c_{2r}^{\dagger}-\frac{13t^{2}}{10\lambda}\sum_{r}c_{2r}^{\dagger}c_{2r}=-\frac{2t^{2}}{5\lambda}-\frac{6t^{2}}{5\lambda}\sum_{r}c_{2r}^{\dagger}c_{2r}, \\
& (H_{\text{eff}})_{22}=-\frac{t^{2}}{10\lambda}\sum_{r}c_{2r}^{\dagger}c_{2r}-\frac{13t^{2}}{10\lambda}\sum_{r}c_{2r}c_{2r}^{\dagger}=-\frac{2t^{2}}{5\lambda}-\frac{6t^{2}}{5\lambda}\sum_{r}c_{2r}c_{2r}^{\dagger}, \\
& (H_{\text{eff}})_{12}=-\mathrm{i}\frac{6t^{2}}{5\lambda}\sum_{r'r}(\sigma_{y}\otimes\mathbb{I})_{r'r}c_{2r'}^{\dagger}c_{2r}^{\dagger}.
\end{align}
The effective Hamiltonian is given by
\begin{align}
H_{\text{eff}}= & \left(\begin{array}{cc}
1 & 0\\
0 & 0
\end{array}\right)(H_{\text{eff}})_{11}+\left(\begin{array}{cc}
0 & 0\\
0 & 1
\end{array}\right)(H_{\text{eff}})_{22}+\left(\begin{array}{cc}
0 & 1\\
0 & 0
\end{array}\right)(H_{\text{eff}})_{12}+\left(\begin{array}{cc}
0 & 0\\
1 & 0
\end{array}\right)[(H_{\text{eff}})_{12}]^{\dagger}\\
= & \frac{1}{2}\mathcal{\mathbb{I}}[(H_{\text{eff}})_{11}+(H_{\text{eff}})_{22}]+\frac{1}{2}S_{z}[(H_{\text{eff}})_{11}-(H_{\text{eff}})_{22}]+\frac{1}{2}S_{x}[(H_{\text{eff}})_{12}+[(H_{\text{eff}})_{12}]^{\dagger}]+\frac{\mathrm{i}}{2}S_{y}[(H_{\text{eff}})_{12}-[(H_{\text{eff}})_{12}]^{\dagger}]\\
= & -\frac{14t^{2}}{5\lambda}\mathcal{\mathbb{I}}+\frac{6t^{2}}{5\lambda}S_{z}J_{z}+\frac{6t^{2}}{5\lambda}S_{x}J_{x}+\frac{6t^{2}}{5\lambda}S_{y}J_{y}.
\end{align}
Based on the form of this effective Hamiltonian, we defined the following
operators.
\begin{equation}
J_{x}=-\frac{\mathrm{i}}{2}\sum_{r'r}(\sigma_{y})_{r'r}[c_{2r'}^{\dagger}c_{2r}^{\dagger}-c_{2r'}c_{2r}];\ J_{y}=\frac{\mathrm{1}}{2}\sum_{r'r}(\sigma_{y})_{r'r}[c_{2r'}^{\dagger}c_{2r}^{\dagger}+c_{2r'}c_{2r}];\ J_{z}=[2-\sum_{r}c_{2r}^{\dagger}c_{2r}].
\end{equation}
It is easy to check that they satisfy the $SU(2)$ Lie algebra, $[J_{z},J_{x}]=2\mathrm{i}J_{y},\ [J_{z},J_{y}]=-2\mathrm{i}J_{x},\ [J_{x},J_{y}]=2\mathrm{i}J_{z}$. As we will see now that $J_{x},J_{y},J_{z}$ are 2-channel electrons if we rewrite them by using spin space (first $2\times 2$ matrices like
$\sigma_{y}$ in $\sigma_{y}\otimes\mathbb{I}$) and channel space
(second $2\times 2$ matrices like $\mathbb{I}$ in $\sigma_{y}\otimes\mathbb{I}$).
$\uparrow,\downarrow$ are usually the label for spin space. $\tau=1,2$
labels the channel space.
\begin{equation}
J_{x}=-\sum_{\tau}(c_{\tau\uparrow}^{\dagger}c_{\tau\downarrow}^{\dagger}-c_{\tau\uparrow}c_{\tau\downarrow}),\ J_{y}=-\mathrm{i}\sum_{\tau}(c_{\tau\uparrow}^{\dagger}c_{\tau\downarrow}^{\dagger}+c_{\tau\uparrow}c_{\tau\downarrow}),\
J_{z}=2-\sum_{\tau}(c_{\tau\uparrow}^{\dagger}c_{\tau\uparrow}+c_{\tau\downarrow}^{\dagger}c_{\tau\downarrow})
\end{equation}
After redefining $f_{\tau}=(f_{\tau\uparrow},f_{\tau\downarrow})^T \equiv (c_{\tau\uparrow}^{\dagger}, c_{\tau\downarrow})^T$, the generators become $J_{x}=\sum_{\tau}f_{\tau}^{\dagger}\sigma_{x}f_{\tau},\ J_{y}=\sum_{\tau}f_{\tau}^{\dagger}\sigma_{y}f_{\tau},\ J_{z}=\sum_{\tau}f_{\tau}^{\dagger}\sigma_{z}f_{\tau}$. Thus, the effective Hamiltonian is actually an isotropic $2$-channel
Kondo interaction:
\begin{equation}
H_{\text{eff}}
=-\frac{14t^{2}}{5\lambda}\mathcal{\mathbb{I}}+\frac{6t^{2}}{5\lambda}\sum_{\tau=1,2}f_{\tau}^{\dagger}(\sigma_{x}S_{x}+\sigma_{y}S_{y}+\sigma_{z}S_{z})f_{\tau}.
\end{equation}
This concludes the derivation of Eq.~\eqref{eq:effecH} of the main text in the case $k =2$.

\subsection{Effective Hamiltonian of $Sp(6)$ Kondo model at strong coupling}

\subsubsection{Representation of the current operator}

Similarly for $Sp(6)$ Kondo model, $1,2$-particle representations of conduction electrons are $J_1^A=T_A$ and $J^{(2)}_A=\langle\beta|J_A|\alpha\rangle=4\Tr(A^{\beta}T_AA^{\alpha})$ with $|\alpha\rangle=\sum_{m_{1}m_{2}}(A^{\alpha})_{m_{1}m_{2}}c_{m_{1}}^{\dagger}c_{m_{2}}^{\dagger}|0\rangle$, $A^{\alpha T}=-A^{\alpha}$, $A^{\alpha\dagger}=A^{\alpha}$ and normalization $2\Tr(A^{\beta}A^{\alpha})=\delta_{\beta\alpha}$. Let us consider the 3-particle representation. Again, we first define the 3-particle. There are two ways
\begin{align}
|\rho\rangle= & \sum_{m_{1}m_{2}m_{3}}(B^{\rho})_{m_{1}m_{2}m_{3}}c_{m_{1}}^{\dagger}c_{m_{2}}^{\dagger}c_{m_{3}}^{\dagger}|0\rangle.\\
|\rho'\rangle= & \sum_{m_{1}'m_{2}'m_{3}'}(B^{\rho'})_{m_{1}'m_{2}'m_{3}'}h_{m_{1}'}^{\dagger}h_{m_{2}'}^{\dagger}h_{m_{3}'}^{\dagger}|6,0\rangle
=  \sum_{m_{1}m_{2}m_{3}}(B^{'\rho'})_{m_{1}m_{2}m_{3}}c_{m_{1}}c_{m_{2}}c_{m_{3}}|6,0\rangle.
\end{align}
We require $B^{\rho}$ to be antisymmetric with respect to $m_{1,2,3}$
and $(B^{'\rho'})_{m_{1}m_{2}m_{3}}=\sum_{m_{1}'m_{2}'m_{3}'}(B^{\rho'})_{m_{1}'m_{2}'m_{3}'}(\mathrm{i}\sigma_{y}\otimes\mathbb{I})_{m'_{1}m_{1}}(\mathrm{i}\sigma_{y}\otimes\mathbb{I})_{m'_{2}m_{2}}(\mathrm{i}\sigma_{y}\otimes\mathbb{I})_{m'_{3}m_{3}}$.
Also, the hole creation operator $h_{m'}^{\dagger}=\sum_{m}(\mathrm{i}\sigma_{y}\otimes\mathbb{I})_{m'm}c_{m}.$
The normalization condition will become
\begin{align}
\langle\sigma|\rho\rangle=6\sum_{m_{1}m_{2}m_{3}}(B^{\sigma})_{m_{1}m_{2}m_{3}}^{*}(B^{\rho})_{m_{1}m_{2}m_{3}}=\delta_{\sigma\rho}, \ \text{and}\ \langle\sigma'|\rho'\rangle=\langle\sigma|\rho\rangle=\delta_{\sigma'\rho'}.
\end{align}
Thus, the 3-particle representation of generators are
\begin{align}
& J^{(3)}_A=\langle\sigma|J_A|\rho\rangle  =18\sum_{n_{1}n_{2}}\sum_{m_{1}m_{2}}(B^{\sigma})_{n_{1}m_{1}m_{2}}^{*}(T_A)_{n_{1}n_{2}}(B^{\rho})_{n_{2}m_{1}m_{2}}.\\
& {J'}^{(3)}_A=\langle\sigma'|J_A|\rho'\rangle 
 18\sum_{m_{1}'m_{1}''m_{2}'m_{3}'}\sum_{m_{1}''}(B^{\sigma'})_{m_{1}''m_{2}'m_{3}'}^{*}(T_A)_{m''_{1}m'_{1}}(B^{\rho'})_{m_{1}'m_{2}'m_{3}'}=J^{(3)}_A.
\end{align}
Let us consider the following $A$ and $B$ tensor
\begin{align}
[A^{\alpha=(\alpha_{1},\alpha_{2})}]_{m_{1}m_{2}} & =-\frac{\mathrm{i}}{2!}\delta_{m_{1}}^{\alpha_{1}}\delta_{m_{2}}^{\alpha_{2}}+\frac{\mathrm{i}}{2!}\delta_{m_{1}}^{\alpha_{2}}\delta_{m_{2}}^{\alpha_{1}}\ \text{with}\ \alpha_{1}<\alpha_{2}.\\{}
[B^{\rho=(\rho_{1},\rho_{2},\rho_{3})}]_{m_{1}m_{2}m_{3}} & =-\frac{\mathrm{i}}{3!}\sum_{\sigma}(-1)^{\text{sign(\ensuremath{\sigma})}}\delta_{m_{1}}^{\rho_{\sigma_{1}}}\delta_{m_{2}}^{\rho_{\sigma_{1}}}\delta_{m_{3}}^{\rho_{\sigma_{3}}}\ \text{with\ }\rho_{1}<\rho_{2}<\rho_{3}.
\end{align}
Here, the sum at the second line above is over all the permutations of $\{1,2,3\}$.
The sign function is $-1$ when the parity of $\sigma$ is odd and
$1$ when the parity of $\sigma$ is even. 

\subsubsection{Strong coupling solution}

Similarly to the case $k =2$, we consider the Kondo energy for $N-$particle sector is simply
\begin{equation}
2 H_K/\lambda=2\sum_{A}T_AJ^{(N)}_A=\sum_{A}[(T_A+J^{(N)}_A)^{2}-(T_A)^{2}-(J^{(N)}_A)^{2}].
\end{equation}
Again, the spectrum follows from the representations at different filling $N$ of the lead site adjacent to the impurities, and the corresponding Casimir operator. The results are shown in table \ref{tab:Sp6}, again $0,\mu_1,\mu_2,\mu_3$ are trivial, fundamental and higher dimensional representations of the symplectic group.
\begin{table}
\begin{centering}
\begin{tabular}{|l|c|c|c|c|c|}
\hline 
$N$& $\mu_{1}\otimes\mu_{1}\ (6\times6)$ & $\sum_{A}(T_A+J^{(1)}_A)^{2}$ & $-\sum_{A}(T_A)^{2}$ & $-\sum_{A}(J^{(1)}_A)^{2}$ & $2 H_K/\lambda$\tabularnewline
\hline 
$1,5$ & $2\mu_{1}\ (21)$  & $8$ & $-7/2$ & $-7/2$ & $1$\tabularnewline
\hline 
$1,5$ &$\mu_{2}\ (14)$  & $6$ & $-7/2$ & $-7/2$ & $-1$\tabularnewline
\hline 
$1,5$ &$0\ (1)$  & $0$ & $-7/2$ & $-7/2$ & $-7$\tabularnewline
\hline 
\hline 
$N$ & $\mu_{1}\otimes[\mu_{2}\oplus0]\ [6\times(14+1)]$ & $\sum_{A}(T_A+J^{(2)}_A)^{2}$ & $-\sum_{A}(T_A)^{2}$ & $-\sum_{A}(J^{(2)}_A)^{2}$ & $2 H_K/\lambda$\tabularnewline
\hline 
$2,4$ & $\mu_{1}+\mu_{2}\ (64)$ & $21/2$ & $-7/2$ & $-6$ & 1\tabularnewline
\hline 
$2,4$ & $\mu_{3}\ (14)$ & $15/2$ & $-7/2$ & $-6$ & -2\tabularnewline
\hline 
$2,4$ & $\mu_{1}\ (6)$ & $7/2$ & $-7/2$ & $-6$ & -6\tabularnewline
\hline 
$2,4$ & $\mu_{1}\ (6)$ & $7/2$ & $-7/2$ & $0$ & 0\tabularnewline
\hline 
\hline 
$N$ & $\mu_{1}\otimes[\mu_{3}\oplus\mu_{1}]\ [6\times(14+6)]$ & $\sum_{A}(T_A+J^{(3)}_A)^{2}$ & $-\sum_{A}(T_A)^{2}$ & $-\sum_{A}(J^{(3)}_A)^{2}$ & $2 H_K/\lambda$\tabularnewline
\hline 
$3$ &$\mu_{1}+\mu_{3}\ (70)$ & $12$ & $-7/2$ & $-15/2$ & $1$\tabularnewline
\hline 
$3$ &$2\mu_{1}\ (21)$ & $8$ & $-7/2$ & $-7/2$ & $1$\tabularnewline
\hline 
$3$ &$\mu_{2}\ (14)$ & $6$ & $-7/2$ & $-15/2$ & $-5$\tabularnewline
\hline 
$3$ &$\mu_{2}\ (14)$ & $6$ & $-7/2$ & $-7/2$ & $-1$\tabularnewline
\hline 
$3$ &$0\ (1)$ & $0$ & $-7/2$ & $-7/2$ & $-7$\tabularnewline
\hline 
\end{tabular}
\par\end{centering}
\caption{Decomposition of $Sp(6)$ Kondo interaction with dimensions and energies for $N$ lead electrons coupled to the impurity site. The numbers in the brackets next to the representations are dimensions. For 2-particle sector, the second $\mu_{1}$ (with $\sum_{A}(J^{(2)}_A)^{2}=0$) is from $\mu_{1}\otimes0$. For 3-particle, $\mu_1+\mu_3$ and the first $\mu_2$ (with $\sum_{A}(J^{(3)}_A)^{2}=15/2$) are from $\mu_{1}\otimes\mu_{3}$. The rest of them (with $\sum_{A}(J^{(3)}_A)^{2}=7/2$) are from $\mu_{1}\otimes\mu_{1}$, which is the same as the 1-particle sector. The sectors with 4 and 5 particles can be obtained by particle-hole conjugation.}
\label{tab:Sp6}
\end{table}
Due to particle-hole symmetry, we can conclude that the lowest energy
is indeed from the singlet states which are at particle number $1,3,5$.
Based on this, odd particle sectors should have the lowest energy for general $Sp(2k)$. We will show the exact expression for all the singlet states of $Sp(2k)$ Kondo model.

\subsubsection{Eigenstates}

Let us take a close look at the states after the fusion with impurity.
The 0-particle sector has states $|N=0,p\rangle=d_{p}^{\dagger}|0\rangle$.
The 1-particle states in natural basis are $|m,p\rangle=c_{m}^{\dagger}d_{p}^{\dagger}|0\rangle$
with $m,p=1\ \text{to}\ 6$. We know that they can be separated by
different Kondo energies (subspaces) from the above analysis. The
dimension of them are $6\times6=21+14+1$. The $21$ states are symmetric
of $m$ and $p$, the rest are antisymmetric matrices but with one
special $\sigma_{y}\otimes\mathbb{I}$ which gives the singlet.
\begin{equation}
|N=1,2\mu_{1},a\rangle=\sum_{mp}(s_A)_{mp}|m,p\rangle;\ |N=1,\mu_{2},b\rangle=\sum_{mp}(t^{b})_{mp}|m,p\rangle;\ |N=1,0\rangle=-\frac{\mathrm{\mathrm{i}}}{\sqrt{6}}\sum_{mp}(\sigma_{y}\otimes\mathbb{I})_{mp}|m,p\rangle.
\end{equation}
Here $s_A(a=1,\cdots,21)$ are symmetric matrices and $t^{b}(b=1,\cdots,14)$
are antisymmetric matrices. Similarly, the 2-particle and 3-particle
states are generally defined as
\begin{align}
|N=2,c\rangle= & \sum_{\alpha p}(r^{c})_{\alpha p}|\alpha,p\rangle=\sum_{\alpha p}(r^{c})_{\alpha p}\sum_{m_{1}m_{2}}(A^{\alpha})_{m_{1}m_{2}}c_{m_{1}}^{\dagger}c_{m_{2}}^{\dagger}d_{p}^{\dagger}|0\rangle.\\
|N=3,d\rangle= & \sum_{\rho p}(u^{d})_{\rho p}|\rho,p\rangle=\sum_{\rho p}(u^{d})_{\rho p}\sum_{m_{1}m_{2}m_{3}}(B^{\rho})_{m_{1}m_{2}m_{3}}c_{m_{1}}^{\dagger}c_{m_{2}}^{\dagger}c_{m_{3}}^{\dagger}d_{p}^{\dagger}|0\rangle\\
= & \sum_{\rho'p}(u^{d})_{\rho'p}|\rho',p\rangle=\sum_{\rho'p}(u^{d})_{\rho'p}\sum_{m_{1}m_{2}m_{3}}(B^{'\rho'})_{m_{1}m_{2}m_{3}}c_{m_{1}}c_{m_{2}}c_{m_{3}}d_{p}^{\dagger}|6,0\rangle.
\end{align}
We can define the four different decompositions with different sets
of $c$.
\begin{subequations}
\begin{align}
& (2H_{K}/\lambda)|N=2,\mu_{1}+\mu_{2},c=1,\cdots,64\rangle=  |N=2,c=1,\cdots,64\rangle;\\
& (2H_{K}/\lambda)|N=2,\mu_{3},c=65,\cdots,78\rangle=  -2|N=2,c=65,\cdots,78\rangle;\\
& (2H_{K}/\lambda)|N=2,\mu_{1},c=79,\cdots,84\rangle=  -6|N=2,c=79,\cdots,84\rangle;\\
& (2H_{K}/\lambda)|N=2,\mu_{1},c=85,\cdots,90\rangle=  0.
\end{align}
and
\begin{align}
& (2H_{K}/\lambda)|N=3,\mu_{1}+\mu_{3},d=1,\cdots,70\rangle=  |N=2,d=1,\cdots,70\rangle;\\
& (2H_{K}/\lambda)|N=3,2\mu_{1},d=71,\cdots,91\rangle=  |N=2,d=71,\cdots,91\rangle;\\
& (2H_{K}/\lambda)|N=3,\mu_{2},d=92,\cdots,105\rangle=  -5|N=2,d=92,\cdots,105\rangle;\\
& (2H_{K}/\lambda)|N=3,\mu_{2},d=106,\cdots,119\rangle=  -|N=2,d=106,\cdots,119\rangle;\\
& (2H_{K}/\lambda)|N=3,0,d=120\rangle=  -7|N=2,d=120\rangle;
\end{align}
\end{subequations}
Here, $|N=3,0,d=120\rangle$ is an singlet which has the lowest energy. It is
\begin{equation}
|N=3,0,d=120\rangle=\frac{1}{24\text{\ensuremath{\sqrt{3}}}}\sum_{m_{1}m_{2}m_{3}pij}\epsilon_{m_{1}m_{2}m_{3}pij}(\sigma_{y}\otimes\mathbb{I})_{ij}c_{m_{1}}^{\dagger}c_{m_{2}}^{\dagger}c_{m_{3}}^{\dagger}d_{p}^{\dagger}|0\rangle
\end{equation}
with normalization $\langle N=3,0,d=120|N=3,0,d=120\rangle=1$. The 4,5 and 6-particle sector are particle-hole symmetric with 2,1,0-particle sector. The 6-particle sector has states $|N=6,p\rangle=d_{p}^{\dagger}|6,0\rangle$.
The 5-particle sector is the same as 1-particle sector except that
all the particle operators should be replaced by hole operators. For
example, the lowest energy states in this sector is
\begin{equation}
|N=5,0\rangle=-\frac{\mathrm{i}}{\sqrt{6}}\sum_{m'p}(\sigma_{y}\otimes\mathbb{I})_{m'p}h_{m'}^{\dagger}d_{p}^{\dagger}|6,0\rangle=\frac{1}{\sqrt{6}}\sum_{mm'p}(\sigma_{y}\otimes\mathbb{I})_{m'p}(\sigma_{y}\otimes\mathbb{I})_{m'm}c_{m}d_{p}^{\dagger}|6,0\rangle=-\frac{1}{\sqrt{6}}\sum_{m}c_{m}d_{m}^{\dagger}|6,0\rangle.
\end{equation}
The states in 4-particle sector should be
\begin{align}
|N=4,c'\rangle= & \sum_{\alpha'p'}(r'^{c'})_{\alpha'p'}\sum_{m_{1}'m_{2}'}(A^{\alpha'})_{m_{1}'m_{2}'}h_{m_{1}'}^{\dagger}h_{m_{2}'}^{\dagger}d_{p'}^{\dagger}|6,0\rangle=\sum_{\alpha'p'}(r'^{c'})_{\alpha p}\sum_{m_{1}m_{2}}(C^{\alpha'})_{m_{1}m_{2}}c_{m_{1}}c_{m_{2}}d_{p'}^{\dagger}|6,0\rangle.
\end{align}
where $C^{\alpha}=(\sigma_{y}\otimes\mathbb{I})A^{\alpha}(\sigma_{y}\otimes\mathbb{I})$
gives the same set of antisymmetric matrices. Similar to 2-particle states, we define the 4-particle states as $|\alpha'\rangle=\sum_{m_{1}m_{2}}(C^{\alpha'})_{m_{1}m_{2}}c_{m_{1}}c_{m_{2}}|6,0\rangle$. The normalization condition can be deduced: $\langle\beta'|\alpha'\rangle=2\Tr(C^{\beta'}C^{\alpha'})=\delta_{\beta'\alpha'}$.
Thus, the 4-particle representation of generators are $J^{(4)}_A=\langle\beta'|J_A|\alpha'\rangle=-4\Tr(C^{\alpha'}T_AC^{\beta'})=J^{(2)}_A$.
After diagonalizing $(T_A+J^{(4)}_A)^{2}$, we found that $(r'^{c'})_{\alpha'p'}=(r^{c'})_{\alpha'p'}$
because $J^{(4)}_A=J^{(2)}_A$. In conclusion, the singlet states at $N=1,3,5$ has the lowest energy.
They are
\begin{align}
 & |N=1,0\rangle=-\frac{\mathrm{\mathrm{i}}}{\sqrt{6}}\sum_{mp}(\sigma_{y}\otimes\mathbb{I})_{mp}|m,p\rangle=\frac{\mathrm{\mathrm{1}}}{8\sqrt{6}}\sum_{mpijkl}\epsilon_{mpijkl}(\sigma_{y}\otimes\mathbb{I})_{ij}(\sigma_{y}\otimes\mathbb{I})^{kl}c_{m}^{\dagger}d_{p}^{\dagger}|0\rangle,\\
 & |N=3,0,d=120\rangle=\frac{1}{24\text{\ensuremath{\sqrt{3}}}}\sum_{m_{1}m_{2}m_{3}pij}\epsilon_{m_{1}m_{2}m_{3}pij}(\sigma_{y}\otimes\mathbb{I})_{ij}c_{m_{1}}^{\dagger}c_{m_{2}}^{\dagger}c_{m_{3}}^{\dagger}d_{p}^{\dagger}|0\rangle,\\
 & |N=5,0\rangle=-\frac{1}{\sqrt{6}}\sum_{p}c_{p}d_{p}^{\dagger}|6,0\rangle=-\frac{1}{5!\sqrt{6}}\sum_{m_{1}m_{2}m_{3}m_{4}m_{5}p}\epsilon_{m_{1}m_{2}m_{3}m_{4}m_{5}p}c_{m_{1}}^{\dagger}c_{m_{2}}^{\dagger}c_{m_{3}}^{\dagger}c_{m_{4}}^{\dagger}c_{m_{5}}^{\dagger}d_{p}^{\dagger}|0\rangle.
\end{align}
The above form for the singlet states can be easily generalized into
$Sp(2k)$.

\subsubsection{Perturbative inclusion of leads}

Now, we can calculate the effective Hamiltonian. The hopping energy
between the first and second site is $H_{12}/t=\sum_{r}c_{1r}^{\dagger}c_{2r}+c_{2r}^{\dagger}c_{1r}$.
\begin{subequations}
\begin{align}
(H_{\text{eff}})_{11}= & \sum_{p'=1}^{6}\frac{|\langle N=0,p'|H_{12}|N=1,0\rangle|^{2}}{(-7\lambda/2-0)}+\sum_{c=1}^{64}\frac{|\langle N=2,c|H_{12}|N=1,0\rangle|^{2}}{(-7\lambda/2-\lambda/2)}+\sum_{c=65}^{78}\frac{|\langle N=2,c|H_{12}|N=1,0\rangle|^{2}}{(-7\lambda/2+2\lambda/2)}\notag \\
 & +\sum_{c=79}^{84}\frac{|\langle N=2,c|H_{12}|N=1,0\rangle|^{2}}{(-7\lambda/2+6\lambda/2)}+\sum_{c=85}^{90}\frac{|\langle N=2,c|H_{12}|N=1,0\rangle|^{2}}{(-7\lambda/2-0)}\notag \\
= & -\frac{t^{2}}{21\lambda}\sum_{r}c_{2r}c_{2r}^{\dagger}-\frac{11t^{2}}{7\lambda}\sum_{r}c_{2r}^{\dagger}c_{2r}, 
\end{align}
\begin{align}
(H_{\text{eff}})_{22}= & \sum_{c=1}^{64}\frac{|\langle N=2,c|H_{12}|N=3,d=120\rangle|^{2}}{(-7\lambda/2-\lambda/2)}+\sum_{c=65}^{78}\frac{|\langle N=2,c|H_{12}|N=3,d=120\rangle|^{2}}{(-7\lambda/2+2\lambda/2)}\notag \\
 & +\sum_{c=79}^{84}\frac{|\langle N=2,c|H_{12}|N=3,d=120\rangle|^{2}}{(-7\lambda/2+6\lambda/2)}+\sum_{c=85}^{90}\frac{|\langle N=2,c|H_{12}|N=3,d=120\rangle|^{2}}{(-7\lambda/2-0)}\notag \\
 & +\sum_{c'=1}^{64}\frac{|\langle N=4,c'|H_{12}|N=3,d=120\rangle|^{2}}{(-7\lambda/2-\lambda/2)}+\sum_{c'=65}^{78}\frac{|\langle N=4,c'|H_{12}|N=3,d=120\rangle|^{2}}{(-7\lambda/2+2\lambda/2)}\notag \\
 & +\sum_{c'=79}^{84}\frac{|\langle N=4,c'|H_{12}|N=3,d=120\rangle|^{2}}{(-7\lambda/2+6\lambda/2)}+\sum_{c'=85}^{90}\frac{|\langle N=4,c'|H_{12}|N=3,d=120\rangle|^{2}}{(-7\lambda/2-0)}\notag \\
= & -\frac{17t^{2}}{21\lambda}\sum_{r}c_{2r}c_{2r}^{\dagger}-\frac{17t^{2}}{21\lambda}\sum_{r}c_{2r}^{\dagger}c_{2r},  \end{align}
 \begin{align}
(H_{\text{eff}})_{33}= & \sum_{p'=1}^{6}\frac{|\langle N=6,p'|H_{12}|N=5,0\rangle|^{2}}{(-7\lambda/2-0)}+\sum_{c'=1}^{64}\frac{|\langle N=4,c'|H_{12}|N=5,0\rangle|^{2}}{(-7\lambda/2-\lambda/2)}+\sum_{c'=65}^{78}\frac{|\langle N=4,c'|H_{12}|N=5,0\rangle|^{2}}{(-7\lambda/2+2\lambda/2)}\notag \\
 & +\sum_{c'=79}^{84}\frac{|\langle N=4,c'|H_{12}|N=5,0\rangle|^{2}}{(-7\lambda/2+6\lambda/2}+\sum_{c'=85}^{90}\frac{|\langle N=4,c'|H_{12}|N=5,0\rangle|^{2}}{(-7\lambda/2-0)}\notag \\
= & -\frac{t^{2}}{21\lambda}\sum_{r}c_{2r}^{\dagger}c_{2r}-\frac{11t^{2}}{7\lambda}\sum_{r}c_{2r}c_{2r}^{\dagger}, 
\end{align}
\begin{align}
(H_{\text{eff}})_{12}=[(H_{\text{eff}})_{21}]^{\dagger}
= & \sum_{c=1}^{64}\frac{\langle N=1,0|H_{12}|N=2,c\rangle \langle N=2,c|H_{12}|N=3,d=120\rangle}{(-7\lambda/2-\lambda/2)}\notag \\
&+\sum_{c=65}^{78}\frac{\langle N=1,0|H_{12}|N=2,c\rangle\langle N=2,c|H_{12}|N=3,d=120\rangle}{(-7\lambda/2+2\lambda/2)}\notag \\
& +\sum_{c=79}^{84}\frac{\langle N=1,0|H_{12}|N=2,c\rangle\langle N=2,c|H_{12}|N=3,d=120\rangle}{(-7\lambda/2+6\lambda/2)}\notag \\
& +\sum_{c=85}^{90}\frac{\langle N=1,0|H_{12}|N=2,c\rangle\langle N=2,c|H_{12}|N=3,d=120\rangle}{(-7\lambda/2-0)}\notag \\
= & -\frac{16\sqrt{2}t^{2}}{21\lambda}\sum_{r'r}(\sigma_{y}\otimes\mathbb{I})_{r'r}c_{2r'}^{\dagger}c_{2r}^{\dagger},
\end{align}
\begin{align}
(H_{\text{eff}})_{23}=[(H_{\text{eff}})_{32}]^{\dagger}= & \sum_{c'=1}^{64}\frac{\langle N=3,d=120|H_{12}|N=4,c'\rangle\langle N=4,c'|H_{12}|N=5,0\rangle}{(-7\lambda/2-\lambda/2)}\notag \\
 & +\sum_{c'=65}^{78}\frac{\langle N=3,d=120|H_{12}|N=4,c'\rangle\langle N=4,c'|H_{12}|N=5,0\rangle}{(-7\lambda/2+2\lambda/2)}\notag \\
 & +\sum_{c'=79}^{84}\frac{\langle N=3,d=120|H_{12}|N=4,c'\rangle\langle N=4,c'|H_{12}|N=5,0\rangle}{(-7\lambda/2+6\lambda/2)}\notag \\
 & +\sum_{c'=85}^{90}\frac{\langle N=3,d=120|H_{12}|N=4,c'\rangle\langle N=4,c'|H_{12}|N=5,0\rangle}{(-7\lambda/2-0)}\notag \\
= & -\frac{16\sqrt{2}t^{2}}{21\lambda}\sum_{r'r}(\sigma_{y}\otimes\mathbb{I})_{r'r}c_{2r'}^{\dagger}c_{2r}^{\dagger}.
\end{align}
\end{subequations}
We used the following equations:
\begin{subequations}
\begin{align}
&\langle N=0,p'|H_{12}|N=1,0\rangle=-\frac{\mathrm{i}t}{\sqrt{6}}\sum_{r}(\sigma_{y}\otimes\mathbb{I})_{rp'}c_{2r}^{\dagger},\ \langle N=1,0|H_{12}|N=2,c\rangle= \frac{-2\mathrm{i}t}{\sqrt{6}}\sum_{r}\sum_{\alpha p}(r^{c})_{\alpha p}[A^{\alpha}(\sigma_{y}\otimes\mathbb{I})]_{rp}c_{2r}^{\dagger}.\\
&\langle N=2,c|H_{12}|N=3,d=120\rangle= 6t\sum_{r}\sum_{\alpha\rho}\sum_{p}(r^{c})_{\alpha p}^{*}(u^{120})_{\rho p}\sum_{m_{2}m_{3}}(A^{\alpha})_{m_{2}m_{3}}^{*}(B^{\rho})_{m_{2}m_{3}r}c_{2r}^{\dagger}.\\
&\langle N=3,d=120|H_{12}|N=4,c'\rangle= 6t\sum_{r}\sum_{\alpha\rho}\sum_{p}(r'^{c'})_{\alpha p}(u^{120})_{\rho p}^{*}\sum_{m_{1}m_{2}}(C^{\alpha})_{m_{1}m_{2}}(D^{\rho})_{m_{1}m_{2}r}^{*}c_{2r}^{\dagger},\\
& \langle N=4,c'|H_{12}|N=5,0\rangle=\frac{2t}{\sqrt{6}}\sum_{r}\sum_{\alpha p}(r'^{c'})_{\alpha p}^{*}(C^{\alpha})_{rp}c_{2r}^{\dagger},\  \langle N=5,0|H_{12}|N=6,p'\rangle=-\frac{t}{\sqrt{6}}c_{2p'}^{\dagger}.
\end{align}
\end{subequations}

where $(D^{\rho'})_{m_{1}m_{2}r}=(1/6)\sum_{m_{1}'m_{2}'m_{3}'}\epsilon_{m_{3}'m_{2}'m_{1}'m_{1}m_{2}r}(B^{\rho'})_{m_{1}'m_{2}'m_{3}'}$. We can check the particle-hole symmetry: $(H_{33})_{p'p}=\sum_{r'r}(\sigma_{y}\otimes\mathbb{I})_{p'r'}(H_{11})_{r'r}(\sigma_{y}\otimes\mathbb{I})_{rp},\
(H_{22})_{p'p}=\sum_{r'r}(\sigma_{y}\otimes\mathbb{I})_{p'r'}(H_{22})_{r'r}(\sigma_{y}\otimes\mathbb{I})_{rp},\
(H_{12})_{p'p}=\sum_{r'r}(\mathrm{i}\sigma_{y}\otimes\mathbb{I})_{p'r'}(H_{32})_{r'r}(\mathrm{i}\sigma_{y}\otimes\mathbb{I})_{rp}$. Finally, we can write
\begin{align}
H_{\text{eff}}= & (H_{\text{eff}})_{11}\left(\begin{array}{ccc}
1 & 0 & 0\\
0 & 0 & 0\\
0 & 0 & 0
\end{array}\right)+(H_{\text{eff}})_{22}\left(\begin{array}{ccc}
0 & 0 & 0\\
0 & 1 & 0\\
0 & 0 & 0
\end{array}\right)+(H_{\text{eff}})_{33}\left(\begin{array}{ccc}
0 & 0 & 0\\
0 & 0 & 0\\
0 & 0 & 1
\end{array}\right) \notag \\
 & +(H_{\text{eff}})_{12}\left(\begin{array}{ccc}
0 & 1 & 0\\
0 & 0 & 0\\
0 & 0 & 0
\end{array}\right)+(H_{\text{eff}})_{23}\left(\begin{array}{ccc}
0 & 0 & 0\\
0 & 0 & 1\\
0 & 0 & 0
\end{array}\right)+[(H_{\text{eff}})_{12}]^{\dagger}\left(\begin{array}{ccc}
0 & 0 & 0\\
1 & 0 & 0\\
0 & 0 & 0
\end{array}\right)+[(H_{\text{eff}})_{23}]^{\dagger}\left(\begin{array}{ccc}
0 & 0 & 0\\
0 & 0 & 0\\
0 & 1 & 0
\end{array}\right) \label{eq:SPSixFinal}
\end{align}
where
\begin{align}
&(H_{\text{eff}})_{11} =-\frac{2 t^{2}}{7\lambda}-\frac{32 t^{2}}{21\lambda}\sum_{r}c_{2r}^{\dagger}c_{2r},\
(H_{\text{eff}})_{22}=-\frac{34t^{2}}{7\lambda},\
(H_{\text{eff}})_{33}=-\frac{2t^{2}}{7\lambda}-\frac{32t^{2}}{21\lambda}\sum_{r}c_{2r}c_{2r}^{\dagger},\\
&(H_{\text{eff}})_{12}=[(H_{\text{eff}})_{21}]^{\dagger}=  -\frac{16\sqrt{2}t^{2}}{21\lambda}\sum_{r'r}(\sigma_{y}\otimes\mathbb{I})_{r'r}c_{2r'}^{\dagger}c_{2r}^{\dagger},\
(H_{\text{eff}})_{23}=[(H_{\text{eff}})_{32}]^{\dagger}=  -\frac{16\sqrt{2}t^{2}}{21\lambda}\sum_{r'r}(\sigma_{y}\otimes\mathbb{I})_{r'r}c_{2r'}^{\dagger}c_{2r}^{\dagger}.
\end{align}
If we introduce spin-$1$ matrices with $S_{z}=\text{diag}(1,0,-1)$
and $S_{z}^{2}=\text{diag}(1,0,1)$, we can write 
\begin{align}
 & H_{\text{eff}}=\frac{1}{2}[(H_{\text{eff}})_{11}+(H_{\text{eff}})_{33}-2(H_{\text{eff}})_{22}]S_{z}^{2}+\frac{1}{2}[(H_{\text{eff}})_{11}-(H_{\text{eff}})_{33}]S_{z}+(H_{\text{eff}})_{22}\mathbb{I}+\frac{1}{\sqrt{2}}(H_{\text{eff}})_{12}S_{+}+\frac{1}{\sqrt{2}}[(H_{\text{eff}})_{12}]^{\dagger}S_{-}\\
= & (H_{\text{eff}})_{22}\mathbb{I}+\frac{1}{2}[(H_{\text{eff}})_{11}-(H_{\text{eff}})_{33}]S_{z}+\frac{1}{\sqrt{2}}[(H_{\text{eff}})_{12}+(H_{\text{eff}})_{12}]^{\dagger}]S_{x}+\frac{\mathrm{i}}{\sqrt{2}}[(H_{\text{eff}})_{12}-(H_{\text{eff}})_{12}]^{\dagger}]S_{y}.
\end{align}
We used here $(H_{\text{eff}})_{12}=(H_{\text{eff}})_{23}$, $(H_{\text{eff}})_{11}+(H_{\text{eff}})_{33}-2(H_{\text{eff}})_{22}=0$
and 
\begin{align}
 & \frac{1}{2}[(H_{\text{eff}})_{11}-(H_{\text{eff}})_{33}]=\frac{16t^{2}}{21\lambda}\sum_{r}c_{2r}c_{2r}^{\dagger}-c_{2r}^{\dagger}c_{2r},\ \frac{1}{\sqrt{2}}[(H_{\text{eff}})_{12}+(H_{\text{eff}})_{12}]^{\dagger}]=-\frac{16t^{2}}{21\lambda}\sum_{r'r}(\sigma_{y}\otimes\mathbb{I})_{r'r}[c_{2r'}^{\dagger}c_{2r}^{\dagger}+c_{2r'}c_{2r}]\\
 & \frac{\mathrm{i}}{\sqrt{2}}[(H_{\text{eff}})_{12}-(H_{\text{eff}})_{12}]^{\dagger}]=-\mathrm{i}\frac{16t^{2}}{21\lambda}\sum_{r'r}(\sigma_{y}\otimes\mathbb{I})_{r'r}[c_{2r'}^{\dagger}c_{2r}^{\dagger}-c_{2r'}c_{2r}],
 \end{align} and the spin-1 matrices
 $ S_{-}\equiv S_{x}-\mathrm{i}S_{y}=\sqrt{2}\left(\begin{array}{ccc}
0 & 0 & 0\\
1 & 0 & 0\\
0 & 1 & 0
\end{array}\right)\,,\quad S_{+}\equiv S_{x}+\mathrm{i}S_{y}=\sqrt{2}\left(\begin{array}{ccc}
0 & 1 & 0\\
0 & 0 & 1\\
0 & 0 & 0
\end{array}\right).
$
Still, we have 
\begin{align}
J_{x}= & -\frac{1}{2}\sum_{r'r}(\sigma_{y}\otimes\mathbb{I})_{r'r}[c_{2r'}^{\dagger}c_{2r}^{\dagger}+c_{2r'}c_{2r}],\ J_{y}=  -\frac{\mathrm{i}}{2}\sum_{r'r}(\sigma_{y}\otimes\mathbb{I})_{r'r}[c_{2r'}^{\dagger}c_{2r}^{\dagger}-c_{2r'}c_{2r}],\ J_{z}=  \frac{1}{2}\sum_{r}c_{2r}c_{2r}^{\dagger}-c_{2r}^{\dagger}c_{2r}.
\end{align}
It is easy to check that they satisfy $[J_{x},J_{y}]=2\mathrm{i}J_{z},\ [J_{y},J_{z}]=2\mathrm{i}J_{x},\ [J_{z},J_{x}]= 2\mathrm{i}J_{y}$. Also, they can be rewritten using chanel index $\tau=1,2,3$
and spin index $\uparrow,\downarrow$:
\begin{align}
J_{x}=\mathrm{i}\sum_{\tau}(c_{\tau\uparrow}^{\dagger}c_{\tau\downarrow}^{\dagger}+c_{\tau\uparrow}c_{\tau\downarrow}),\
J_{y}=\sum_{\tau}(-c_{\tau\uparrow}^{\dagger}c_{\tau\downarrow}^{\dagger}+c_{\tau\uparrow}c_{\tau\downarrow}),\ J_{z}= 3-\sum_{\tau}(c_{\tau\uparrow}^{\dagger}c_{\tau\uparrow}+c_{\tau\downarrow}^{\dagger}c_{\tau\downarrow}).
\end{align}
We again define $f_{\tau}=(f_{\tau\uparrow},f_{\tau\downarrow})^{T}$
with $f_{\tau\uparrow}=c_{\tau\uparrow}^{\dagger}$, $f_{\tau\downarrow}=\mathrm{i}c_{\tau\downarrow}$
and $\tau=1,2,3$. Then, we get $J_{x}= \sum_{\tau}f_{\tau}^{\dagger}\sigma_{x}f_{\tau},\
J_{y}=\sum_{\tau}f_{\tau}^{\dagger}\sigma_{y}f_{\tau},\
J_{z}= \sum_{\tau}f_{\tau}^{\dagger}\sigma_{z}f_{\tau}$. Thus, we find the effective Kondo Hamiltonian 
\begin{align}
H_{{\rm eff}} & =-\frac{34t^{2}}{7\lambda}\mathbb{I}_{3\times3}+\frac{32t^{2}}{21\lambda}\sum_{\tau=1}^{3}f_{\tau}^{\dagger}(\sigma_{x}S_{x}+\sigma_{y}S_{y}+\sigma_{z}S_{z})f_{\tau}
\end{align}
Hence, we conclude that $Sp(6)$ Kondo model is equivalent to spin-$1$
$3$-channel Kondo model because $(S_{x},S_{y},S_{z})$ are spin-1 operators. This concludes the derivation of Eq.~\eqref{eq:effecH} of the main text in the case $k = 3$.

\subsection{Strong coupling ground states at arbitrary $k$.}

Let us consider the general $Sp(2k)$ with singlet states appearing
at $1,3,\cdots,2k-1$-particle sector. The unnormalized singlet states
are
\begin{equation}
|N=2j-1\rangle=\sum_{m_{1}\cdots m_{2j-1}pn_{1}\cdots n_{2k-2j}}\epsilon_{m_{1}\cdots m_{2j-1}pn_{1}\cdots n_{2k-2j}}[(\sigma_{y})_{n_{1}n_{2}}\cdots(\sigma_{y})_{n_{2k-2j-1}n_{2k-2j}}]c_{m_{1}}^{\dagger}c_{m_{2}}^{\dagger}\cdots c_{m_{2j-1}}^{\dagger}d_{p}^{\dagger}|0\rangle \label{eq:singletstats}
\end{equation}
where $j=1,\cdots,k$. We first confirm that these states have dimensionless energy 
$-(2k+1)/2$
by acting on them with $\sum_{A}S_AJ_A$,
\begin{align}
 & \sum_{A}S_AJ_A|N=2j-1\rangle=-\frac{(2j-1)}{2}\sum_{m_{1}\cdots m_{2j-1}pn_{1}\cdots n_{2k-2j}}\epsilon_{m_{1}\cdots m_{2j-1}pn_{1}\cdots n_{2k-2j}} [(\sigma_{y})_{n_{1}n_{2}}\cdots(\sigma_{y})_{n_{2k-2j-1}n_{2k-2j}}]\notag\\
 & \times c_{m_{1}}^{\dagger}c_{m_{2}}^{\dagger}\cdots c_{m_{2j-1}}^{\dagger}d_{p}^{\dagger}|0\rangle-\frac{(2j-1)}{2}\sum_{m_{1}\cdots m_{2j-1}pn_{1}\cdots n_{2k-2j}} \epsilon_{m_{2}\cdots m_{2j-1}m_{1}pn_{1}\cdots n_{2k-2j}}(\sigma_{y})_{m_{1}p}(\sigma_{y})_{n_{1}n_{2}}\cdots(\sigma_{y})_{n_{2k-2j-1}n_{2k-2j}}\notag\\
 & \times c_{m_{2}}^{\dagger}\cdots c_{m_{2j-1}}^{\dagger}\sum_{si}(\sigma_{y})_{is}c_{s}^{\dagger}d_{i}^{\dagger}|0\rangle.
\end{align}
Here we used $\sum_{A}\tau^{sm_{1}}_A\tau^{ip}_A=[\delta_{sp}\delta_{im_{1}}-(\sigma_{y})_{is}(\sigma_{y})_{m_{1}p}]/2$. The first term above gives $-[(2j-1)/2]|N=2j-1\rangle$ and the second term is $-[(2k-2j+2)/2]|N=2j-1\rangle$ because of
\begin{align}
 & \sum_{m_{1}\cdots m_{2j-1}pn_{1}\cdots n_{2k-2j}}(\sigma_{y})_{m_{1}p}(\sigma_{y})_{n_{1}n_{2}}\cdots(\sigma_{y})_{n_{2k-2j-1}n_{2k-2j}}[(\sigma_{y})_{is}\epsilon_{m_{2}\cdots m_{2j-1}m_{1}pn_{1}\cdots n_{2k-2j}}\notag\\
 & -(\sigma_{y})_{im_{2}}\epsilon_{sm_{3}\cdots m_{2j-1}m_{1}pn_{1}\cdots n_{2k-2j}}-\cdots-(\sigma_{y})_{im_{2j-1}}\epsilon_{m_{2}\cdots sm_{1}pn_{1}\cdots n_{2k-2j}}]\notag\\
= & -(2k-2j+2)\sum_{m_{2}\cdots m_{2j-1}isn_{1}\cdots n_{2k-2j}}\epsilon_{m_{2}\cdots m_{2j-1}isn_{1}\cdots n_{2k-2j}}(\sigma_{y})_{n_{1}n_{2}}\cdots(\sigma_{y})_{n_{2k-2j-1}n_{2k-2j}}.
\end{align}
Finally, we confirm that 
\begin{equation}
\sum_{A}S_AJ_A|N=2j-1\rangle=\Big[-\frac{2j-1}{2}-\frac{(2k-2j+2)}{2}\Big]|N=2j-1\rangle=-\frac{2k+1}{2}|N=2j-1\rangle.
\end{equation}
The defined states $|N=2j-1\rangle$ are really the ground (singlet) states.

\subsection{Susceptibility at strong coupling}

The $Sp(2k)$ magnetization is determined by applying Zeeman fields to the topological edge states on the island~\cite{KoenigTsvelik2022}. We consider the ground state overlaps of $S_z^{(j)} = d^\dagger_j \sigma_z d_j$ for given $j$ at strong coupling. For example in the case $k = 2$ the two ground states are given by Eqs.~\eqref{eq:N1},\eqref{eq:N3} of the main text.
Acting with the magnetization we obtain
\begin{subequations}
\begin{align}
d^\dagger_j \sigma_z d_j\ket{N = 1}_{\rm singlet} &= -i d^\dagger_j \sigma_z\sigma_y c^*_j \ket{0}_c \otimes \ket{BCS}_d,\\
d^\dagger_j \sigma_z d_j\ket{N = 3}_{\rm singlet} &= d^\dagger_j \sigma_z c_j \ket{4}_c \otimes \ket{BCS}_d.
\end{align}
\end{subequations}
So we readily see that $\langle{N = 1\vert S_z^{(j)} \vert N = 1}\rangle = \langle{N = 1\vert S_z^{(j)} \vert N = 3}\rangle = \langle{N = 3\vert S_z^{(j)} \vert N =3}\rangle = 0$. As a consequence the $S_z^{(j)}$ susceptibility (which is the susceptibility of the symplectic ``Zeeman'' fields) involves excited states at the strong coupling solution and thus will be suppressed as opposed to the $SU(2)_k$ susceptibilty, which leads to additional suppression in the scaling exponents. Similar arguments hold for $k>2$.  Therefore, the dimension of the leading irrelevant operator in the conventional k-channel $SU(2)$ does not determine the scaling of $Sp(2k)$ susceptibility, which instead is Fermi liquid like to leading order.

\section{Derivation of conductances }
\label{app:conductance}
In this section, we evaluate the zero-temperature conductance at the free and Kondo fixed points, following closely Ref.~\cite{Oshikawa_2006}. 
We start from the free fixed point, Sec.~\ref{sec:Gfree}, where the Kondo coupling is zero, $\lambda =0$, and we find a vanishing conductance as expected.  
In Sec.~\ref{sec:GKondo} we derive the conductance in the Kondo (intermediate coupling) fixed point. 
Finally, in Sec.~\ref{sec:Gweak} we derive the conductance in the tunneling limit, near the weak-coupling $Sp(2k)$ fixed point. 

\subsection{Free fixed point conductance \label{sec:Gfree}}
We can linearize the kinetic energy term (without boundary) of this $Sp(2k)$ model near the Fermi energy: $\epsilon_{p}-\epsilon_{F}=\tau\hbar v_{F}p$
where $\tau=R(+),L(-)$. For simplicity, we take $\epsilon_{F}=0$.
Thus, the linearized free Hamiltonian is
\begin{equation}
H_{0}=\sum_{\alpha=1}^{2k}\sum_{\tau}\sum_{p}\tau\hbar v_{F}p\,c_{\tau,p,\alpha}^{\dagger}c_{\tau,p,\alpha}.
\end{equation}
The current for flavor $\rho$ is defined as
\begin{align}
I_{\tau,\rho}(x)=  ev_{F}\psi_{\tau,\rho}^{\dagger}(x)\psi_{\tau,\rho}(x)=\frac{ev_{F}}{L}\sum_{k}\mathrm{e}^{\mathrm{i}kx}n_{\tau,k,\rho}
\end{align}
where $\psi_{\tau,\rho}(x)=\frac{1}{\sqrt{L}}\sum_{p}\mathrm{e}^{\mathrm{i}px}c_{\tau,p,\rho}$
and we defined that $n_{\tau,k,\rho}=\sum_{p'}c_{\tau,p',\rho}^{\dagger}c_{\tau,p'+k,\rho}$. It is known that $[H_{0},n_{\tau,k,\rho}]= -\tau\hbar v_{F}kn_{\tau,k,\rho}$.
Thus, by applying this commutation relation,
we get 
\begin{equation}
I_{\tau,\rho}(x,t)=\mathrm{e}^{\mathrm{i}H_{0}t/\hbar}I_{\tau,\rho}(x)\mathrm{e}^{-\mathrm{i}H_{0}t/\hbar}=\frac{ev_{F}}{L}\sum_{k}\mathrm{e}^{\mathrm{i}kx}\mathrm{e}^{\mathrm{i}H_{0}t/\hbar}n_{\tau,k,\rho}\mathrm{e}^{-\mathrm{i}H_{0}t/\hbar}=\frac{ev_{F}}{L}\sum_{k}n_{\tau,k,\rho}\mathrm{e}^{\mathrm{i}k(x-\tau v_{F}t)}.
\end{equation}
The correlation functions will be
\begin{equation}
\langle[I_{\tau,\rho}(y,t),I_{\tau',\rho'}(x,0)]\rangle_{\text{bulk}}=\mathrm{i}\frac{e^{2}v_{F}^{2}}{(2\pi)^{2}}\tau\delta_{\tau\tau'}\delta_{\rho\rho'}\lim_{\eta\to0^{+}}\frac{4\eta(y-x-\tau v_{F}t)}{[\eta^{2}+(y-x-\tau v_{F}t)^{2}]^{2}}.
\end{equation}
where we used the anomalous commutator $[n_{\tau,k,\rho},n_{\tau',k',\rho'}]=\frac{kL\tau}{2\pi}\delta_{k,-k'}\delta_{\tau\tau'}\delta_{\rho\rho'}$ and regularized the following integral by taking $z_{\tau}=y-x-\tau v_{F}t$
and
\begin{equation}
\int_{-\infty}^{\infty}dk\,k\,\mathrm{e}^{\mathrm{i}kz_{\tau}}= \lim_{\eta\to0^{+}}\int_{-\infty}^{0}dk\,k\,\mathrm{e}^{\mathrm{i}k(z_{\tau}-\mathrm{i}\eta)}+\lim_{\eta\to0^{+}}\int_{0}^{\infty}dk\,k\,\mathrm{e}^{\mathrm{i}k(z_{\tau}+\mathrm{i}\eta)}=\lim_{\eta\to0^{+}}\mathrm{i}\frac{4\eta z_{\tau}}{(\eta^{2}+z_{\tau}^{2})^{2}}.
\end{equation}
If we consider a boundary with the left and right chiral
fields like $\psi_{L,\rho}(x,t)=S\psi_{R,\rho}(-x,t)$ where $S$
is only a phase factor, this leads $\psi_{L,\rho}(x,t)^{\dagger}\psi_{L,\rho}(x,t)=\psi_{R,\rho}^{\dagger}(-x,t)\psi_{R,j,\rho}(-x,t)$ which is $I_{L,\rho}(x,t)=I_{R,\rho}(-x,t)$. Thus, the above correlation functions become
\begin{align}
 & \langle[I_{L,\rho}(y,t),I_{R,\rho'}(x,0)]\rangle_{\text{free}}=\langle[I_{R,\rho}(-y,t),I_{R,\rho'}(x,0)]\rangle_{\text{bulk}}\notag \\
= & -\mathrm{i}\frac{e^{2}v_{F}^{2}}{(2\pi)^{2}}\delta_{\rho\rho'}\lim_{\eta\to0^{+}}\frac{4\eta(y+x+v_{F}t)}{[\eta^{2}+(y+x+v_{F}t)^{2}]^{2}}=\langle[I_{L,\rho}(y+x,t),I_{L,\rho'}(0,0)]\rangle_{\text{bulk}}.
\end{align}
\begin{align}
 & \langle[I_{R,\rho}(y,t),I_{L,\rho'}(x,0)]\rangle_{\text{free}}=\langle[I_{R,\rho}(y,t),I_{R,\rho'}(-x,0)]\rangle_{\text{bulk}}\notag \\
= & \mathrm{i}\frac{e^{2}v_{F}^{2}}{(2\pi)^{2}}\delta_{\rho\rho'}\lim_{\eta\to0^{+}}\frac{4\eta(y+x-v_{F}t)}{[\eta^{2}+(y+x-v_{F}t)^{2}]^{2}}=\langle[I_{R,\rho}(y+x,t),I_{R,\rho'}(0,0)]\rangle_{\text{bulk}}.
\end{align}

The conductance can be calculated by the above correlation functions:
\begin{equation}
G_{\rho\rho'}=\Re\lim_{\omega\to0}\frac{1}{\hbar\omega}\int_{0}^{\infty}\mathrm{d}t\,\mathrm{e}^{\mathrm{i}\omega t}\frac{1}{L}\int_{0}^{L}\mathrm{d}x\,\langle[I_{\rho}(y,t),I_{\rho'}(x,0)]\rangle_{\text{free}}
\end{equation}
with $y>0$ for simplicity and $I_{\rho}(y,t)=I_{R,\rho}(y,t)-I_{L,\rho}(y,t)$. It gives
\begin{align}
G_{\rho\rho'}
= & \Re\lim_{\omega\to0}\frac{1}{\hbar\omega L}\int_{0}^{L}\mathrm{d}x\int_{0}^{\infty}\mathrm{d}t\,\mathrm{e}^{\mathrm{i}\omega t}\langle[I_{R,\rho}(y-x,t),I_{R,\rho'}(0,0)]-[I_{L,\rho}(y+x,t),I_{L,\rho'}(0,0)]\notag \\
 & -[I_{R,\rho}(y+x,t),I_{R,\rho'}(0,0)]+[I_{L,\rho}(y-x,t),I_{L,\rho'}(0,0)]\rangle_{\text{free}}.
\end{align}
As we have calculated, 
\begin{align}
 & \langle[I_{R,\rho}(u,t),I_{R,\rho'}(0,0)]\rangle_{\text{bulk}}=\frac{e^{2}v_{F}^{2}}{(2\pi)^{2}}\delta_{\rho\rho'}\lim_{\eta\to0^{+}}\mathrm{i}\frac{4\eta(u-v_{F}t)}{[\eta^{2}+(u-v_{F}t)^{2}]^{2}},\\
 & \langle[I_{L,\rho}(u,t),I_{L,\rho'}(0,0)]\rangle_{\text{bulk}}=\frac{e^{2}v_{F}^{2}}{(2\pi)^{2}}\delta_{\rho\rho'}\lim_{\eta\to0^{+}}\mathrm{i}\frac{4\eta(-u-v_{F}t)}{[\eta^{2}+(u+v_{F}t)^{2}]^{2}}.
\end{align}
We determine the following integral for calculating the conductance by taking $t\to s+u/v_{F}$ and $\lim_{\eta\to0^{+}}-\mathrm{i}\frac{4\eta v_{F}s}{[\eta^{2}+(v_{F}s)^{2}]^{2}}=\mathrm{i}\frac{2}{v_{F}}\partial_{s}\lim_{\eta\to0^{+}}\frac{\eta}{\eta^{2}+(v_{F}s)^{2}}=\mathrm{i}\frac{2\pi}{v_{F}^{2}}\partial_{s}\delta(s)$:
\begin{align}
 f(u)=\Re\int_{0}^{\infty}\mathrm{d}t\,\mathrm{e}^{\mathrm{i}\omega t}\lim_{\eta\to0^{+}}\mathrm{i}\frac{4\eta(u-v_{F}t)}{[\eta^{2}+(u-v_{F}t)^{2}]^{2}}=\frac{2\pi}{v_{F}^{2}}\omega H(u)\cos(\omega u/v_{F})
\end{align}
where $H(u)$ is the Heaviside step function. Finally,
\begin{equation}
G_{\rho\rho'}
=  \frac{e^{2}}{h}\delta_{\rho\rho'}\frac{1}{L}\int_{0}^{L}\mathrm{d}x\,[H(y-x)-H(-y-x)-H(y+x)+H(-y+x)].
\end{equation}
Because $x,y>0$, we get $G_{\rho\rho'}=0$ in the free fixed point (no Kondo interaction).

At the same time, the $L,R$
conductance with $\rho=\rho'$ (because it is diagonal in the flavor space, we call it $G_{L,R;11}^{\text{free}}$) is
\begin{align}
G_{L,R;11}^{\text{free}}= & \Re\lim_{\omega\to0}\frac{1}{\hbar\omega L}\int_{0}^{L}\mathrm{d}x\int_{0}^{\infty}\mathrm{d}t\,\mathrm{e}^{\mathrm{i}\omega t}\langle[I_{R,\rho}(y,t),I_{L,\rho}(x,0)]\rangle_{\text{free}} \nonumber \\
= & \Re\lim_{\omega\to0}\frac{1}{\hbar\omega L}\int_{0}^{L}\mathrm{d}x\int_{0}^{\infty}\mathrm{d}t\,\mathrm{e}^{\mathrm{i}\omega t}\langle[I_{R,\rho}(y+x,t),I_{R,\rho}(0,0)]\rangle_{\text{free}}
=  \frac{e^{2}}{h}\frac{1}{L}\int_{0}^{L}\mathrm{d}x\,H(y+x)=\frac{e^{2}}{h}\equiv G_{0}.
\end{align}
Here, we used the commutator correlation functions to calculate the conductance which is equivalent to using time-ordered correlation functions with time integral from $-\infty$ to $\infty$. From the CFT method, Ref.~\cite{affleck1995conformal}, we know the correlation function $\langle I_{L,1}(y,t)I_{R,2}(x,0)\rangle_{\text{Kondo}}$
is proportional to $\langle I_{L,1}(y,t)I_{R,1}(x,0)\rangle_{\text{free}}$ with an overall constant factor, which also applies to the above current commutator correlation functions. From here on, we are interested in the ratios of correlation functions and for brevity will suppress writing the commutator and the arguments for the operators. Hence, it implies that we can compare the correlation functions at critical Kondo coupling with the nonzero $L,R$ correlation functions at free case to get the conductance:
\begin{equation}
G_{L,R;12}^{\text{Kondo}}=\frac{\langle I_{L,1} I_{R,2}\rangle_{\text{Kondo}}}{\langle I_{L,1} I_{R,1}\rangle_{\text{free}}} G_{L,R;11}^{\text{free}} = \frac{\langle I_{L,1} I_{R,2}\rangle_{\text{Kondo}}}{\langle I_{L,1} I_{R,1}\rangle_{\text{free}}} G_0.
\end{equation}

\subsection{Kondo fixed point conductance \label{sec:GKondo}}
We will get the overall factors between certain density correlation functions to further calculate the conductance at the Kondo coupling fixed point. The charge current and spin current are
\begin{equation}
I_{X,j}^{c}=I_{X,j,\uparrow}+I_{X,j,\downarrow};\ I_{X,j}^{s}=I_{X,j,\uparrow}-I_{X,j,\downarrow}.
\end{equation}
Depending on the symmetry of the (Kondo) Hamiltonian, we impose the following relation
\begin{equation}
\langle I_{L,j}^{c}I_{R,j}^{c}\rangle=\langle I_{L,1}^{c}I_{R,1}^{c}\rangle,\,\langle I_{L,j}^{c}I_{R,k}^{c}\rangle=\langle I_{L,1}^{c}I_{R,2}^{c}\rangle\,(j\neq k);\ \langle I_{L,j}^{s}I_{R,j}^{s}\rangle=\langle I_{L,1}^{s}I_{R,1}^{s}\rangle,\,\langle I_{L,j}^{s}I_{R,k}^{s}\rangle=\langle I_{L,1}^{s}I_{R,2}^{s}\rangle\,(j\neq k).
\end{equation}
The charge density $J_{X}^{c}$ , spin density $J_{X}^{s}$, flavor
density $J_{X}^{f}$ and the spin-flavor density $J_{X}^{sf}$ 
\begin{equation}
J_{X}^{c}=\sum_{j=1}^{k}I_{X,j}^{c};\ J_{X}^{s}=\sum_{j=1}^{k}I_{X,j}^{s};\ J_{X}^{f}=I_{X,1}^{c}-I_{X,2}^{c};\ J_{X}^{sf}=I_{X,1}^{s}-I_{X,2}^{s}.
\end{equation}
Thus, the correlation function of these two densities are
\begin{subequations}
\begin{align}
&\langle J_{L}^{c}J_{R}^{c}\rangle=  \sum_{i=1}^{k}\sum_{j=1}^{k}\langle I_{X,i}^{c}I_{X,j}^{c}\rangle=k\langle I_{L,1}^{c}I_{R,1}^{c}\rangle+k(k-1)\langle I_{L,1}^{c}I_{R,2}^{c}\rangle;\\
&\langle J_{L}^{s}J_{R}^{s}\rangle=  \sum_{i=1}^{k}\sum_{j=1}^{k}\langle I_{X,i}^{s}I_{X,j}^{s}\rangle=k\langle I_{L,1}^{s}I_{R,1}^{s}\rangle+k(k-1)\langle I_{L,1}^{s}I_{R,2}^{s}\rangle;\\
&\langle J_{L}^{f}J_{R}^{f}\rangle=\langle(I_{L,1}^{c}-I_{L,2}^{c})(I_{R,1}^{c}-I_{R,2}^{c})\rangle=2\langle I_{L,1}^{c}I_{R,1}^{c}\rangle-2\langle I_{L,1}^{c}I_{R,2}^{c}\rangle;\\
&\langle J_{L}^{sf}J_{R}^{sf}\rangle=\langle(I_{L,1}^{s}-I_{L,2}^{s})(I_{R,1}^{s}-I_{R,2}^{s})\rangle=2\langle I_{L,1}^{s}I_{R,1}^{s}\rangle-2\langle I_{L,1}^{s}I_{R,2}^{s}\rangle.
\end{align}
\end{subequations}
\begin{subequations}
Inversely,
\begin{align} 
& \langle I_{L,1}^{c}I_{R,1}^{c}\rangle=\frac{1}{k^{2}}\langle J_{L}^{c}J_{R}^{c}\rangle+\frac{k-1}{2k}\langle J_{L}^{f}J_{R}^{f}\rangle;\
  \langle I_{L,1}^{c}I_{R,2}^{c}\rangle=\frac{1}{k^{2}}\langle J_{L}^{c}J_{R}^{c}\rangle-\frac{1}{2k}\langle J_{L}^{f}J_{R}^{f}\rangle;\\
& \langle I_{L,1}^{s}I_{R,1}^{s}\rangle=\frac{1}{k^{2}}\langle J_{L}^{s}J_{R}^{s}\rangle+\frac{k-1}{2k}\langle J_{L}^{sf}J_{R}^{sf}\rangle;\
  \langle I_{L,1}^{s}I_{R,2}^{s}\rangle=\frac{1}{k^{2}}\langle J_{L}^{s}J_{R}^{s}\rangle-\frac{1}{2k}\langle J_{L}^{sf}J_{R}^{sf}\rangle.
\end{align}
\end{subequations}

In the free case, only diagonal current correlation functions will survive, which means $\langle I_{L,1}^{c,s}I_{R,2}^{c,s}\rangle_{\text{f}}=0$. When the Kondo interaction is tunned on at the fixed point, the off-diagonal terms appear. In order to calculate them, we relate them to the density correlation functions, each of which leads to an overall factor between the density correlation functions with Kondo interaction at the fixed point and the density correlation functions at the free case (without Kondo interaction). We define the factors in the following way:
\begin{equation}
\langle J_{L}^{c}J_{R}^{c}\rangle_{\text{K}}=\langle J_{L}^{c}J_{R}^{c}\rangle_{\text{f}};\ \langle J_{L}^{s}J_{R}^{s}\rangle_{\text{K}}=u_{s}\langle J_{L}^{s}J_{R}^{s}\rangle_{\text{f}};\ \langle J_{L}^{f}J_{R}^{f}\rangle_{\text{K}}=u_{f}\langle J_{L}^{f}J_{R}^{f}\rangle_{\text{f}};\ \langle J_{L}^{sf}J_{R}^{sf}\rangle_{\text{K}}=u_{sf}\langle J_{L}^{sf}J_{R}^{sf}\rangle_{\text{f}}.
\end{equation}
They give
\begin{subequations}
\begin{align}
&\langle I_{L,1}^{c}I_{R,1}^{c}\rangle_{\text{K}}=\frac{1+(k-1)u_{f}}{k}\langle I_{L,1}^{c}I_{R,1}^{c}\rangle_{\text{f}};\ \langle I_{L,1}^{c}I_{R,2}^{c}\rangle_{\text{K}}=\frac{1-u_{f}}{k}\langle I_{L,1}^{c}I_{R,1}^{c}\rangle_{\text{f}}. \\
&\langle I_{L,1}^{s}I_{R,1}^{s}\rangle_{\text{K}}=\frac{u_{s}+(k-1)u_{sf}}{k}\langle I_{L,1}^{s}I_{R,1}^{s}\rangle_{\text{f}};\ \langle I_{L,1}^{s}I_{R,2}^{s}\rangle_{\text{K}}=\frac{u_{s}-u_{sf}}{k}\langle I_{L,1}^{s}I_{R,1}^{s}\rangle_{\text{f}}.
\end{align}
\end{subequations}
The charge conductance is equal to
\begin{equation}
\frac{G_{12}}{G_{0}}=\frac{\langle I_{L,1}^{c}I_{R,2}^{c}\rangle_{\text{K}}}{\langle I_{L,1}^{c}I_{R,1}^{c}\rangle_{\text{f}}}=\frac{1-u_{f}}{k}.
\end{equation}

Next, we determine the representations that the four densities belong
to. The four densities defined above can be rewritten as
\begin{equation}
J_{X}^{c}=\psi_{X}^{\dagger}[\mathbb{I}_{k\times k}\otimes\mathbb{I}_{2\times2}]\psi_{X};\ J_{X}^{s}=\psi_{X}^{\dagger}[\mathbb{I}_{k\times k}\otimes\sigma_{z}]\psi_{X};\ J_{X}^{f}=\psi_{X}^{\dagger}[D_{1}\otimes\mathbb{I}_{2\times2}]\psi_{X};\ J_{X}^{sf}=\psi_{X}^{\dagger}[D_{1}\otimes\sigma_{z}]\psi_{X}
\end{equation}
where $D_{1}=\text{diag}(1,-1,0,\cdots,0)$. Under $Sp(2k)$ transformations, the currents transform as $J'=SJS^{\dagger}$
where $S\in Sp(2k)$. $J_{X}^{c}$ should be an singlet ($0$) since
it transforms to itself under $Sp(2k)$ transformation (forming a
1-dimensional subspace). $J_{X}^{s}$ and $J_{X}^{sf}$ belong to
the adjoint representation ($2\mu_{1}$) with dimension $k(2k+1)$
because their intermediate matrices are the generators. $J_{X}^{f}$
belong to a $(k-1)(2k+1)$-dimensional representation ($\mu_{2}$).
These three representations are all the representations constructed
by $\psi^{\dagger}M\psi$ with $M$ a $2k\times2k$ Hermitian matrix.
The number of independent $2k\times2k$ Hermitian matrix is $(2k)^{2}=1+k(2k+1)+(k-1)(2k+1)$
which are the dimensions of the three representations. The singlet
is simply the non-zero trace term. The adjoint representation is given
by $[M(\sigma_{y}\otimes\mathbb{I})]^{T}=M(\sigma_{y}\otimes\mathbb{I})$
and traceless (automatically satisfied by the first constrain). The
last one is also given by traceless but $[M(\sigma_{y}\otimes\mathbb{I})]^{T}=-M(\sigma_{y}\otimes\mathbb{I})$.
In terms of Dynkin lables, their representations are correspondingly
$(0,\cdots,0),\,(2,0,\cdots,0)$ and $(0,1,0,\cdots,0)$ . 

It is known that the above $u_{1,2,3}$ are calculated through modular $S$-matrix
of $Sp(2k)_{1}$ which is (given by Ref.~\cite{hung2018linking})
\begin{equation}
S_{R_{1}R_{2}}=(-1)^{k(k-1)/2}\left(\frac{2}{k+2}\right)^{k/2}\det[G_{R_{1}R_{2}}],\ G_{R_{1}R_{2};i,j}=\sin\left[\frac{\pi\phi_{R_{1};i}\phi_{R_{2};j}}{k+2}\right]
\end{equation}
where $R_{1,2}=(a_{1},\cdots,a_{k})$ are Dynkin labels with $a_{i}$
integer, $\phi_{R;i}=l_{i}-i+k+1,\,i=1,\cdots,k$ and $ l_{i}=\sum_{n=i}^{k}a_{n}$.
By doing the algebra, we get
\begin{equation}
u_{s}=u_{sf}=\frac{S_{\mu_{1}}^{2\mu_{1}}/S_{0}^{2\mu_{1}}}{S_{\mu_{1}}^{0}/S_{0}^{0}}=1-\frac{2\sin^{2}\left[\frac{\pi}{k+2}\right]}{\cos\left[\frac{\pi}{k+2}\right]},\ u_{f}=\frac{S_{\mu_{1}}^{\mu_{2}}/S_{0}^{\mu_{2}}}{S_{\mu_{1}}^{0}/S_{0}^{0}}=\frac{\cos\left[\frac{3\pi}{k+2}\right]}{\cos\left[\frac{\pi}{k+2}\right]}. 
\end{equation}
The charge conductance is now
\begin{equation}
\frac{G_{12}}{G_{0}}=\frac{1-u_{f}}{k}=\frac{4\sin^{2}\left[\frac{\pi}{k+2}\right]}{k}.
\end{equation} 
This result is similar to the conductance found for the   $k$-channel spin-$1/2$ $SU(2)$  charge Kondo effect~\cite{PhysRevB.57.R5579,PhysRevB.65.195101}. 
There, the fixed point off-diagonal conductance is proportional to $1-S_{k}$ where $S_{k} =   4\cos^2 \frac{\pi}{k+2} - 3$  is the relevant ratio [identical to $u_f$ above] of modular $S$-matrices for $SU(2)_{k}$. 

\subsection{Weak coupling conductance}\label{sec:Gweak}
In this section, we derive the conductance matrix in the limit of weak coupling, where $\lambda$ in Eq.~(\ref{eq:HKondo}) of the main text is small. 
In the tunneling limit the currents are weak and it is convenient to use the Kubo formula in the form, 
\begin{equation}
    G_{ij} =  \Re\lim_{\omega\to0}\frac{\mathrm{i}}{\hbar\omega}\lim_{\mathrm{i}\omega_{n}\to\omega+\mathrm{i}0^{+}}\int_{0}^{\hbar\beta}\mathrm{\mathrm{d}}\tau\mathrm{e}^{\mathrm{i}\omega_{n}\tau}\langle T_{\tau}I_{i}(\tau)I_{j}(0)\rangle_{0}, \label{eq:GKubo}
\end{equation}
written  in the imaginary-time formalism. 

We first start with the spin-resolved current operators $I_{\rho}$ where $\rho=1\! \uparrow,\dots,k\! \uparrow,1 \! \downarrow,\dots,k\! \downarrow$. 
To find the current operators $I_{\rho}$, we note that in the full Hamiltonian $H=H_{0}+H_{K}$, the kinetic term $H_{0}$ conserves the particle number $N_{\rho} = \int_{-\infty}^{\infty}\mathrm{d}x\,\psi_{\rho}^{\dagger}(x)\psi_{\rho}(x)$. 
The current is defined by its time derivative:
\begin{equation}
I_{\rho}=-e\frac{\mathrm{d}N_{\rho}}{\mathrm{d}t}=-\mathrm{i}\frac{e}{\hbar}[H,N_{\rho}],\ 
\end{equation} 
which is non-zero due to $H_K$. 
The $Sp(2k)$ Kondo interaction is
\begin{equation}
H_{K}=\sum_{A\alpha\beta}\lambda_{A}S^{A}\psi_{\alpha}^{\dagger}(0)(T^{A})_{\alpha\beta}\psi_{\beta}(0).
\end{equation}
Now, we can calculate the conductance by considering weak coupling $\lambda_A$ through current-current correlation functions. Before that, we fist derive the current defined above by the commutator $[H,N_{\rho}]=[H_K,N_{\rho}]$. Thus, we get
\begin{align}
I_{\rho}(0) & =-\mathrm{i}\frac{e}{\hbar}[H,N_{\rho}]=-\mathrm{i}\frac{e}{\hbar}\sum_{A}\lambda_{A}S^{A}\sum_{\beta}\left[(T^{A})_{\beta\rho}\psi_{\beta}^{\dagger}(0)\psi_{\rho}(0)-(T^{A})_{\rho\beta}\psi_{\rho}^{\dagger}(0)\psi_{\beta}(0)\right].
\end{align}
If we introduce the imaginary time $\tau=\mathrm{i}t$, the time-dependent
current is $I_{\rho}(\tau)=\mathrm{e}^{H\tau}I_{\rho}(0)\mathrm{e}^{-H\tau}$.
We can calculate the following average
\begin{align}
\langle T_{\tau}I_{\rho}(\tau)I_{\sigma}(0)\rangle_{0}
= \frac{e^{2}}{\hbar^{2}}G(-\tau)G(\tau)\sum_{AB}\lambda_{A}\lambda_{B}\langle S^{A}S^{B}\rangle\Big[(T^{A})_{\sigma\rho}(T^{B})_{\rho\sigma}+(T^{A})_{\rho\sigma}(T^{B})_{\sigma\rho}-(T^{B}T^{A})_{\sigma\rho}\delta_{\rho\sigma}-(T^{A}T^{B})_{\rho\sigma}\delta_{\rho\sigma}\Big]
\end{align}
where we defined the Green's functions
\begin{equation}
\langle T_{\tau}\psi_{i,\alpha}(\tau)\psi_{j,\beta}^{\dagger}(0)\rangle_{0}=G(\tau)\delta_{ij}\delta_{\alpha\beta};\ \langle T_{\tau}\psi_{i,\alpha}^{\dagger}(\tau)\psi_{j,\beta}(0)\rangle_{0}=-G(-\tau)\delta_{ij}\delta_{\alpha\beta}.
\end{equation}
Denoting $\rho_{0}$ the density of states per spin, per length at the Fermi level, we find 
\begin{align}
G(\tau)= & \frac{1}{L}\sum_{p}\mathrm{e}^{-\epsilon_{p}\tau/\hbar}[\theta(\tau)[1-f(\epsilon_{p})]-\theta(-\tau)f(\epsilon_{p})]=\lim_{\eta\to0^{+}}\rho_{0}\frac{\pi/\beta}{\sin[\pi(\tau+\text{sgn(\ensuremath{\tau})}\eta)/\beta\hbar]}.
\end{align}
We simplify the above correlation functions by taking $\langle S^{A}S^{B}\rangle=C_{S}\delta^{AB}$ where $C_{S}=\Tr(T^{A})^{2}=1$, which leads to
\begin{equation}
\langle T_{\tau}I_{\rho}(\tau)I_{\sigma}(0)\rangle_{0}=\frac{2e^{2}}{\hbar^{2}}G(-\tau)G(\tau)\sum_{A}\lambda_{A}^{2}\left\{ |(T^{A})_{\sigma\rho}|^{2}-[(T^{A})^{2}]_{\sigma\rho}\delta_{\rho\sigma}\right\} .
\end{equation}
This current-current correlation function is still between spin-resolved currents, with $\rho$ and $\sigma$ taking $2k$ values. 

To obtain the charge current for channel $i=1,\dots,k$ we sum over the spin indices by taking 
\begin{equation}
I_{i}(\tau)=\sum_{\rho}R_{i\rho}I_{\rho}(\tau),\ R 
=\left(\begin{array}{cc}
\mathbb{I}_{k\times k} & \mathbb{I}_{k\times k} 
\end{array}\right).
\end{equation}
The charge current correlation function between different channels $i,j$ is thus
\begin{equation}
\langle T_{\tau}I_{i}(\tau)I_{j}(0)\rangle_{0}=\sum_{\rho\sigma}R_{i\rho}R_{j\sigma}\langle T_{\tau}I_{\rho}(\tau)I_{\sigma}(0)\rangle_{0}=\frac{2e^{2}}{\hbar^{2}}G(-\tau)G(\tau)\sum_{A}\lambda_{A}^{2}\sum_{\rho\sigma}R_{i\rho}R_{j\sigma}\left\{ |(T^{A})_{\sigma\rho}|^{2}-[(T^{A})^{2}]_{\sigma\rho}\delta_{\rho\sigma}\right\} .
\end{equation}
Finally, by using Eq.~(\ref{eq:GKubo}),  the conductance is
\begin{align}
G_{ij} 
=  \frac{e^{2}}{h}4\pi^{2}\rho_{0}^{2}\sum_{A}\lambda_{A}^{2}\sum_{\rho\sigma}R_{i\rho}R_{j\sigma}\left\{ |(T^{A})_{\sigma\rho}|^{2}-[(T^{A})^{2}]_{\sigma\rho}\delta_{\rho\sigma}\right\} 
\end{align}
where we used for small $\omega$, 
\begin{equation}
\text{Im}\lim_{\mathrm{i}\omega_{n}\to\omega+\mathrm{i}0^{+}}\int_{0}^{\hbar\beta}\mathrm{\mathrm{d}}\tau\mathrm{e}^{\mathrm{i}\omega_{n}\tau}G(-\tau)G(\tau)=- \pi \hbar^{2}\rho_{0}^{2} \omega.
\end{equation}
In particular, for the isotropic case $\lambda_{A}=\lambda$ with $k=2,3$, we get
\begin{equation}
G_{i\neq j}^{k=2}=\frac{8e^{2}}{h}(\pi\rho_0\lambda)^2, \quad 
G_{i\neq j}^{k=3}=\frac{8e^{2}}{h}(\pi\rho_0\lambda)^2. 
\end{equation}
The diagonal elements are determined by current conservation,  $\sum_{i} G_{ij}=0$.

\end{widetext}

\end{document}